\documentclass[aps,prd,twocolumn,superscriptaddress]{revtex4-2}

\usepackage{graphicx}
\usepackage{subfigure, psfrag}

\usepackage{amsmath}
\usepackage{amssymb}
\usepackage{sistyle}
\usepackage{color}
\usepackage[normalem]{ulem}

\newenvironment{rcases}
  {\left.\begin{aligned}}
  {\end{aligned}\right\rbrace}

\begin{document}

\title{Peculiarities of the Landau level collapse in graphene ribbons
in crossed magnetic and in-plane electric fields}

\author{A.A.~Herasymchuk}
\affiliation{Department of Physics, Taras Shevchenko National University of Kyiv,
64/13, Volodymyrska Street, Kyiv 01601, Ukraine}

\author{S.G.~Sharapov}
\affiliation{Bogolyubov Institute for Theoretical Physics, National Academy of Sciences of Ukraine, 14-b Metrologichna Street, Kyiv 03143 Ukraine}
\affiliation{
Kyiv Academic University, Kyiv 03142, Ukraine}

\author{V.P.~Gusynin}
\affiliation{Bogolyubov Institute for Theoretical Physics, National Academy of Sciences of Ukraine, 14-b Metrologichna Street, Kyiv, 03143, Ukraine}

\date{\today}

\begin{abstract}
Employing the low-energy effective theory alongside a combination of analytical
and numerical techniques, we explore the Landau level collapse phenomenon,
uncovering previously undisclosed features.
We consider both finite-width graphene ribbons and semi-infinite geometries
subjected to a perpendicular magnetic field and an in-plane electric field,
applied perpendicular to both zigzag and armchair edges.
In the semi-infinite geometry  the hole
(electron)-like Landau levels collapse as the ratio of electric and magnetic
fields reaches the critical value $ +(-) 1$.
On the other hand, the energies of the electron (hole)-like levels
remain distinct near the edge and deeply within the bulk approaching
each other asymptotically for the same critical value.
In the finite geometry, we show
that the electron (hole)-like levels become denser and merge,
forming a band.
\end{abstract}


\maketitle

\section{Introduction}

Rabi \cite{Rabi1928ZP} was the one who solved the just-discovered
Dirac equation for a free electron in a homogeneous magnetic field employing
the symmetric gauge.
Four months earlier, Fock \cite{Fock1928ZP} calculated the energy levels of a nonrelativistic electron subjected
to both a magnetic field and a harmonic oscillator potential. However, he did not explore the limit in the absence of the potential,
while the study of Rabi demonstrated the quantization of energy for free electrons.
Frenkel and Bronstein \cite{Frenkel1930} found quantized levels in the magnetic field
now known as Landau levels independently of Landau himself  \cite{Landau1930}.
In fact, the aim of a paper \cite{Frenkel1930} was to
investigate whether the discrete set of energy levels of free electrons
constituted one of the paradoxes associated with the Dirac equation or
corresponded to the real physical phenomenon which is not yet observed
experimentally.

However, the experimental exploration of relativistic-like Landau levels,
distinct from nonrelativistic counterparts,
became attainable almost 80 years later in
condensed-matter systems, thanks to the groundbreaking
discovery of graphene in 2005 \cite{Geim2005,Kim2005}.

Furthermore, the fact that graphene is a two-dimensional
material allows to access the regime when the confining
potential at the edges of graphene nanoribbons is
atomically sharp.
The quantum Hall edge states in this case are defined by
boundary conditions of vanishing electron wave functions at the
crystal edges.

It is worth noting that conventional experiments conducted on two-dimensional semiconductors primarily access the regime characterized by electrostatically reconstructed edges.
In this case, the system reduces its energy by reconfiguring the edge states into steps, which give rise to alternating compressible and incompressible stripes \cite{Chklovskii1992PRB}. Additionally, visualizing these edge states is difficult
because they are buried inside the semiconductors.

Graphene, therefore, offers an opportunity to investigate the
real-space structure of edge states using scanning probe techniques
\cite{Li2012NatCom,Coissard2022}, while avoiding their electrostatic
reconstruction.
Other techniques of visualization of charge
transport through Landau levels are also available
\cite{Nazin2010NatPhys,Tetienne2017ASc,Kim2021NarCom}.

Undoubtedly, the most captivating features arise from the relativistic Landau levels, which lack counterparts in standard electron systems.
One notable phenomenon among them is the Landau level collapse in Ref.~\cite{Lukose2007PRL}
(see also Ref.~\cite{Peres2007JPCM}), and subsequently observed experimentally
in Refs.~\cite{Singh2009PRB,Gu2011PRL}.

For the massive Dirac with the dispersion
$\mathcal{E}(p) = \pm \sqrt{v_F^2 p^2 + \Delta^2}$, where
$v_F$ is the Fermi velocity, $\Delta$ is the gap,
the spectrum in the
perpendicularly crossed magnetic $H$ and electric $E$ fields
reads
\cite{Alisultanov2014PB}
\begin{equation}
\label{LL-collapse}
\begin{split}
\mathcal{E}_n & = \mathcal{E}^\ast_n - \hbar k \frac{c E}{H}, \\
\mathcal{E}^\ast_n & = \pm (1 - \beta^2 )^{3/4}\sqrt{\frac{ 2n \hbar v_F^2 e H}{c}
+ \frac{\Delta^2}{(1 - \beta^2 )^{1/2}}},
\end{split}
\end{equation}
where $n=0,1, \ldots$,
$k$ is the in-plane wave vector along the direction perpendicular to the electric field,
$\beta = v_0/v_F = c E/(v_F H)$.
Here and in what follows we assume that $H > 0$
and use CGS units.

As the dimensionless parameter $\beta$
reaches its critical value, $|\beta_c | =1$,  the Landau level staircase merges
into one level \cite{Lukose2007PRL,Peres2007JPCM}. This collapse of the Landau levels
can be regarded as a transition from the closed elliptic quasiparticle  orbits for
$|\beta| <1$ ($|v_0| < v_F$) to open hyperbolic orbits for $|\beta| >1$ ($|v_0| > v_F$) \cite{Shytov2009SSC}.

For $\Delta =0$, the spectrum Eqs.~(\ref{LL-collapse}) reduces to the spectrum
obtained in \cite{Lukose2007PRL,Peres2007JPCM}.
The generalization for a finite $\Delta$ case was done in
Ref.~\cite{Alisultanov2014PB} (see also recent studies \cite{Arjona20017PRB,Nimyi2022PRB}).

The validity of long wavelength approximation for $\beta \neq 0$
was verified in Ref.~\cite{Lukose2007PRL} by performing numerical
computations using the tight-binding model for graphene lattices of a finite
size with the zigzag edges.
It is stated in Ref.~\cite{Lukose2007PRL} that the Landau level
collapse still occurs at the lower value of $\beta_c \simeq 0.9$.
However, a careful examination of the corresponding figure from Ref.~\cite{Lukose2007PRL} reveals that this phenomenon does not manifest as a collapse in the same manner as in the case of an infinite system. Instead, it signifies an increase in the level density.
As the levels approach one another, their finite width causes them to begin overlapping. In this context, the Landau level collapse does occur on the ribbon; however, its interpretation differs from that of the infinite system case.

Now we briefly overview the relevant literature.
The evolution of edge states in the presence of an electric field
was also explored by numerical computations conducted on a finite lattice
in Refs.~\cite{Roslyak2010PTRS,Ostahie2015PRB}.

Besides the analytical studies of Landau levels in crossed fields on an infinite plane \cite{Lukose2007PRL,Peres2007JPCM}, the levels have also been explored for ribbons and semi-infinite geometries using the low-energy model without an electric field
\cite{Brey2006bPRB,Ababnin2006PRL,Romanosvsky2011PRB,Wang2011EPJB,Gusynin2008PRB,Gusynin2008FNT,Gusynin2009PRB,Deplace2010PRB}.

To the best of our knowledge, the only analytic study of Landau levels
in crossed magnetic and electric fields applied to ribbons was done
in the recent paper of the authors \cite{Herasymchuk2023PSS}.
A special attention was paid to the analytical analysis of dispersionless surface modes localized at the zigzag edge.

The aim of this study is to complement the analysis presented in Ref.~\cite{Herasymchuk2023PSS} by uncovering the undisclosed features of Landau level collapse in half-planes and ribbons with
zigzag edges. Additionally, we broaden our investigation to include ribbons and half-planes with armchair edges. We point out that the very meaning of the
Landau level collapse becomes different in the restricted geometry.

The paper is organized as follows.
In Sec.~\ref{sec:model}, we introduce a low-energy model for a graphene ribbon
with  zigzag and armchair edges
subject to crossed magnetic and
electric fields and represent the main equations in the unified form.
General solutions of these equations in terms of the
parabolic cylinder function are presented in Sec.~\ref{sec:general-solutions}.
The numerical and supporting analytical results in the half-plane geometry for the zigzag and armchair edges are described in Secs.~\ref{sec:zigzag-semi}
(the details of the calculation are provided in
Appendices~\ref{sec:Appendix-probability} and \ref{sec:Appendix-zigzag})
and \ref{sec:armchair-semi}, respectively.
A property of the Landau level
collapse, viz. that some levels do not collapse at the edge
is considered in  Secs.~\ref{sec:zigzag-finiteE-edge}
and \ref{sec:armchar-finiteE-edge}.
The critical regime, $|\beta| =1$, is addressed in
Sec.~\ref{sec:critical}, where the main results of this study are obtained.
In Sec.~\ref{sec:conclusion} the summary of the obtained results is given.
Moreover, the Supplemental Material (SM) \cite{SM} presents
additional calculations useful for comparison with the results given
in the main text.

\section{Model}
\label{sec:model}

To determine eigenenergy $\mathcal{E}$
we consider the stationary Dirac equation,
$H \Psi( \mathbf{r}) = \mathcal{E} \Psi( \mathbf{r})$,
with the Hamiltonian describing  low-energy excitations in graphene,
\begin{equation}
\label{Dirac-E}
H = \hbar
v_{F}\left(-\alpha_{1}iD_{x}-\alpha_{2}iD_{y}\right) + \Delta \alpha_{3} + V(\mathbf{r}) .
\end{equation}
Here the $4\times 4$ $\alpha$ matrices $\alpha_{i}= \tau_3 \otimes \sigma_i$
and the Pauli matrices $\tau_i$, $\sigma_i$ (as well as the $2\times 2$ unit
matrices $\tau_0$, $\sigma_0$) act on the valley ($\mathbf{K}_\eta $ with
$\eta = \pm$)
and sublattice ($A,B$) indices, respectively, of the
four-component spinors $\Psi^T = \left( \Psi_+^T, \Psi_-^T \right) =
\left( \psi_{AK_+}, \psi_{BK_+}, \psi_{BK_-}, \psi_{AK_-} \right)$.
This representation is derived from a tight-binding model for graphene (see, e.g., Ref.~\cite{Gusynin2007review}) and thus allows for the formulation of appropriate boundary conditions for armchair and zigzag edges in the continuum model.
We investigate both massless Dirac-Weyl fermions in pristine graphene and massive Dirac fermions with a mass (gap) parameterized as $\Delta$

The gap $\Delta$ considered in the present paper corresponds to
the time-reversal symmetry conserving gap $\tilde \Delta$
in the notation of \cite{Gusynin2008PRB,Gusynin2008FNT,Gusynin2009PRB}
and is related to the carrier density imbalance between the A and B sublattices.
Recall that this gap can be introduced in graphene
when it is placed on top of hexagonal boron nitride (G/hBN) and
the crystallographic axes of graphene and hBN are aligned.

The  orbital effect of a perpendicular magnetic field $\mathbf{H} = \nabla \times \mathbf{A}$
is included via the covariant spatial
derivative  $D_j=\partial_j+(ie/\hbar c)A_j$ with $j=x,y$ and $-e<0$,
while the potential $V(\mathbf{r})$ corresponds to the
static electric field $e \mathbf{E} =\nabla V(\mathbf{r}) $.
The Zeeman interaction is neglected in this paper, because
of its smallness for moderate values of magnetic field
(see, e.g., Ref.~\cite{Gusynin2007review}).

\begin{figure}[h]
\includegraphics[width=.51\textwidth]{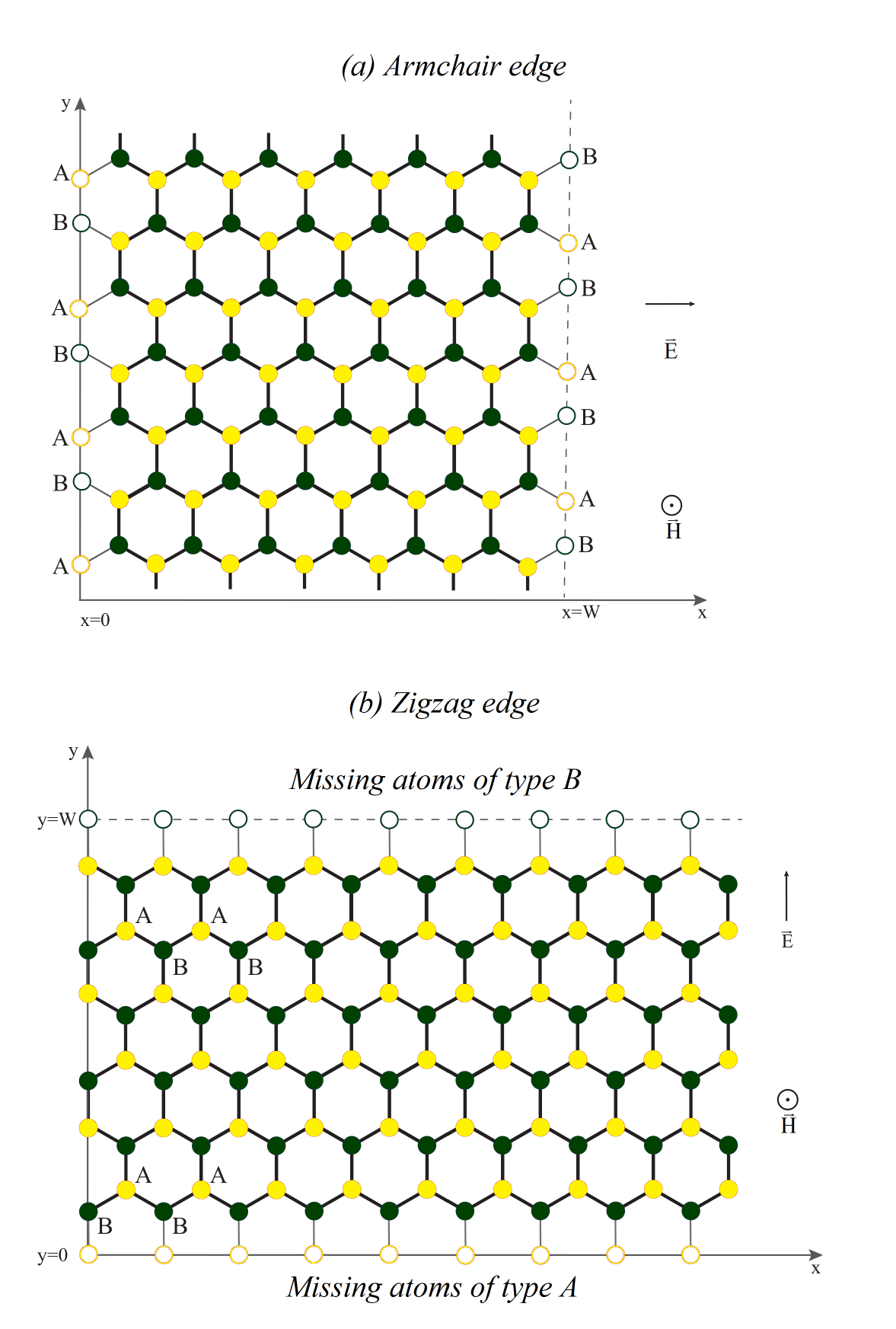}
\caption{
The lattice structure of a finite-width graphene ribbon with
(a) armchair and (b) zigzag edges.}
\label{fig:1}
\end{figure}

We consider the ribbons with the armchair and zigzag edges
as shown in Figs.~\ref{fig:1}~(a) and (b), respectively.
The ribbons are subjected to a combination of crossed uniform magnetic and electric fields. The magnetic field $\mathbf{H}$ is applied perpendicular to the plane of the graphene ribbon along the positive $z$ axis, while the in-plane electric field $\mathbf{E}$ is applied perpendicular to the ribbon edges.

As we will discuss below, only the boundary conditions for the armchair edges
(\ref{boundary-armchair})
involve an admixture of the wave-functions from  both $\mathbf{K}_{\eta}$ points,
while Eq.~(\ref{Dirac-E}) splits into a pair of two independent
Dirac equations for each $\mathbf{K}_{\eta}$ point,
$H_\eta \Psi_\eta( \mathbf{r}) = \mathcal{E} \Psi_\eta( \mathbf{r})$
with the Hamiltonian
\begin{equation}
\label{Dirac-eq-2*2}
H_\eta =
- i  \hbar v_F \eta (\sigma_1 D_x + \sigma_2 D_y) + \eta \Delta \sigma_3 +
V(\mathbf{r}) .
\end{equation}
One can see from Eq.~(\ref{Dirac-eq-2*2}) that having the solutions for $\mathbf{K}_+$
point, the corresponding solutions for $\mathbf{K}_-$ can be obtained by
changing the signs of energy $\mathcal{E}$ and electric field in $V(\mathbf{r})$.
Finally, one should take into account that for the spinor $\Psi_-$ the
components of the spinor corresponding to $A$ and $B$ sublattices are
exchanged as compared to $\Psi_+$.

\subsection{Armchair edge}
\label{sec:armchair}

An armchair edge is parallel to the $y$ as shown in Fig.~\ref{fig:1}~(a).
The in-plane electric field $\mathbf{E}$ is applied respectively in
$x$ direction, so the potential $V(\mathbf{r}) = e E x$.
In this case, it is convenient to consider
the magnetic field $H$ in the following
Landau gauge $(A_{x},A_{y})=(0,H x)$,
where $H$ is the magnitude of a constant
magnetic field orthogonal to the graphene plane.

Accordingly, the differential equations in
Eq.~(\ref{Dirac-E}) do not depend explicitly on the $y$ coordinate.
Therefore, the wave functions are plane waves in the $y$ direction,
\begin{equation}
\label{y-armchair}
\negthickspace \negthickspace
\psi_{\scriptscriptstyle{AK_{\pm}}}(\mathbf{r},k)
=\frac{e^{iky}}{\sqrt{2\pi l}}u_{\scriptscriptstyle{\pm}}(x,k), \thickspace
\psi_{\scriptscriptstyle{BK_{\pm}}} (\mathbf{r},k)
=\frac{e^{iky}}{\sqrt{2\pi l}} v_{\scriptscriptstyle{\pm}}(x,k),
\end{equation}
where $l = \sqrt{\hbar c/ |e H|}$ is the magnetic length.
The wave vector $k$ measures the displacement from $\mathbf{K}_{\eta}$ points.
A specific choice of the coordinate system in Ref.~\cite{Gusynin2007review} defines $\mathbf{K}{\pm} = \pm (2 \pi/a) , (2/3,0)$, where $a$ is the lattice constant. The maximum value of the wave vector $k$ is constrained by the boundaries of the first Brillouin zone.

Recall that the wave vector $k$ determines the center of the electron orbital along the $x$ direction, given by $x_k= -kl^2$. For a system with a ribbon of finite width $W$, such as $0 \leq x \leq W$, the condition that the peak of the wave function lies within the ribbon is met only for eigenstates with wave vectors $k$ in the finite range
$- W/l^2 \leq k \leq 0$. This phenomenon is known as the position-wave vector duality in the Landau gauge.

Substituting Eq.~(\ref{y-armchair}) in Eq.~(\ref{Dirac-E}) we
obtain the following system of equations for $\mathbf{K}_{+}$
point
\begin{equation}
\label{armchair-1}
	\begin{pmatrix}
	\frac{e E x- \mathcal{E} + \Delta}{\hbar v_F} & -i \partial_x - i k - i \frac{e}{\hbar c} H x \\
	-i \partial_x + i k + i \frac{e}{\hbar c} H x & \frac{e E x- \mathcal{E} -\Delta}{\hbar v_F}
	\end{pmatrix} \psi_+ = 0,
	\end{equation}
where the spinor ${\psi}_+^T = (u_+, v_+)$.
One can see that the envelope functions
$u_{+}(x,k)$ and $v_{+}(x,k)$
[$u_{-}(x,k)$ and $v_{-}(x,k)$] depend
only on a single dimensionless combination of the variables,
$\xi=x/l + kl$,
so Eq.~(\ref{armchair-1}) acquires the form
\begin{equation}
\label{Dirac-armchair-dimensionless}
	\begin{pmatrix}
	\beta \xi-\varepsilon + \delta   &  -i\partial_{\xi} - i \xi \\
	-i\partial_{\xi} + i \xi & \beta \xi -\varepsilon - \delta
	\end{pmatrix} \psi_+ = 0.
\end{equation}
Here we introduced the notations for dimensionless quantities
\begin{equation}
\label{dimensionless}
\beta = \frac{c E}{v_F H}, \quad	
\varepsilon =  \frac{l \mathcal{E}}{\hbar v_F} +   \beta k l, \qquad
\delta = \frac{l \Delta}{\hbar v_F} .	
\end{equation}
The important dimensionless parameter $\beta$ in Eq.~(\ref{dimensionless})
describes the strength of the electric field relative to the
magnetic field.
In this paper, we restrict our analysis to the case where $|\beta| \leq 1$ and do not consider the pair creation regime.

To obtain  the energy spectrum we need to supplement
the differential equations for the envelope functions $u_{\pm}(x,k)$ and $v_{\pm}(x,k)$ functions with suitable boundary conditions.
Such conditions can be derived from the tight-binding
model \cite{McCann2004JPCM,Brey2006aPRB,Brey2006bPRB,Ababnin2006PRL}.
Note that in the tight-binding calculation
the values of the total wave vector projected
on the armchair edge direction coincide
for the different $\mathbf{K}_{\eta}$ valleys
(see e.g., Ref.~\cite{Deplace2010PRB}).
This leads to valley admixing by the boundary condition:
\begin{subequations}
\label{boundary-armchair}
\begin{align}
\label{boundaryA-0}
x   = 0:  \qquad \quad & u_{+}(kl)  + u_{-}(kl) =0, \nonumber \\
                       & v_{+}(kl)  + v_{-}(kl) =0,  \\
\label{boundaryA-W}
x   = W:  \qquad \quad & u_{+}(W/l+kl) + u_{-}(W/l+kl) =0, \nonumber \\
                       & v_{+}(W/l+kl) + v_{-}(W/l+kl) =0,
\end{align}
\end{subequations}
see Fig.~\ref{fig:1}.
As we shall see in Sec.~\ref{sec:armchair-semi},
this choice of boundary conditions proves particularly convenient when
transitioning from the ribbon configuration to the semi-infinite,
half-plane geometry with $W \to \infty$.
To consider the symmetry properties of the corresponding solutions, it is
more convenient to choose the ribbon centered at $y=0$, i.e.,
$-W/2 \leq x \leq W/2$.
This case is briefly discussed in Sec.~S3
within the SM \cite{SM}  in parallel to the consideration
made for the zigzag edge case in Sec.~S2 within the SM.

\subsection{Zigzag edge}
\label{sec:zigzag}

A zigzag edge is parallel to the $x$ direction as shown in Fig.~\ref{fig:1}~(b).
The in-plane  electric field $\mathbf{E}$ is applied in
$y$ direction, so the potential $V(\mathbf{r}) = e E y$.
In this case, it is convenient to consider
the magnetic field $H$ in the following
Landau gauge, $(A_{x},A_{y})=(-Hy,0)$.

Accordingly, the differential equations in
Eq.~(\ref{Dirac-E}) do not depend explicitly on the $x$ coordinate.
Therefore, the wave functions are plane waves in the $x$ direction,
\begin{equation}
\label{x-zigzag}
\negthickspace \negthickspace \psi_{\scriptscriptstyle{AK_{\pm}}}(\mathbf{r},k)=\frac{e^{-ikx}}
{\sqrt{2\pi l}}u_{\scriptscriptstyle{\pm}}(y,k), \thickspace
\psi_{\scriptscriptstyle{BK_{\pm}}}(\mathbf{r},k)=\frac{e^{-ikx}}{\sqrt{2\pi
l}}v_{\scriptscriptstyle{\pm}}(y,k).
\end{equation}
The  center of the electron orbital along the $y$
direction is $y_k= - kl^2$. Since
$0 \leq y \leq W$, the wave vectors $k$ are within the range $ - W/l^2 \leq k \leq 0$. Note that the values of the total wave vector for the different
$\mathbf{K}_{\eta}$ valleys in the tight-binding calculation fall
in the different wave vectors domains, because $K_{+x}\neq K_{-x}$.

Substituting Eq.~(\ref{x-zigzag}) in Eq.~(\ref{Dirac-eq-2*2})
we obtain the following system of equations for $\mathbf{K}_{+}$
point
\begin{equation}
\label{zigzag-1}
	\begin{pmatrix}
	\frac{e E y- \mathcal{E} + \Delta}{\hbar v_F} & - \partial_y
-  k - \frac{e}{\hbar c} H y  \\
	 \partial_y -  k -  \frac{e}{\hbar c} H y  & \frac{e E y- \mathcal{E} -\Delta}{\hbar v_F}
	\end{pmatrix} \psi_+ = 0,
	\end{equation}
where $\psi_+^T = (u_+,v_+)$.
One can see that
the envelope functions $u_{+}(y,k)$ and $v_{+}(y,k)$
[$u_{-}(y,k)$ and $v_{-}(y,k)$] depend
only on a single dimensionless combination of the variables,
$\xi=y/l + kl$. Finally,  one can rewrite
Eq.~(\ref{zigzag-1}) in exactly the same form
as Eq.~(\ref{Dirac-armchair-dimensionless}), but for
the  spinor ${\tilde \psi}_+^T = (u_+, -i v_+)$
(see Ref.~\cite{Herasymchuk2023PSS}).
The notation ${\tilde \psi}$,
together with the opposite sign in $\exp (-ikx)$ as compared to
\cite{Gusynin2008PRB,Gusynin2008FNT,Gusynin2009PRB}, allows us to
unify the equations describing zigzag and armchair edges.

To obtain  the energy spectrum we need to supplement
the differential equations for the envelope functions $u_{\pm}(y,k)$ and $v_{\pm}(y,k)$ functions with suitable boundary conditions.
Such conditions can be derived from the tight-binding
model \cite{McCann2004JPCM,Brey2006aPRB,Brey2006bPRB,Ababnin2006PRL}.

In the case of a graphene ribbon of a finite width in the $y$
direction, $0 \leq y \leq W$, and with two zigzag edges parallel to
the $x$ direction, the $A$ and $B$ components of wave functions
should vanish on the opposite edges,
\begin{subequations}
\label{boundary-zigzag}
\begin{align}
\label{boundaryZ-0}
y  & = 0:  \qquad \quad u_{+}(kl)  = u_{-}(kl) =0, \\
\label{boundaryZ-W}
y  & = W:  \qquad v_{+}(W/l+kl)  = v_{-}(W/l+kl) =0,
\end{align}
\end{subequations}
see Fig.~\ref{fig:1}. As for the armchair edges,
this choice of boundary conditions proves particularly convenient when
transition is carried out from the ribbon configuration to the semi-infinite,
half-plane geometry, with $W \to \infty$.
To consider the symmetry properties of the corresponding solutions, it is
more convenient to choose the ribbon centered at $y=0$, i.e.,
$-W/2 \leq y \leq W/2$. This case is studied in Sec.~S2 within the SM \cite{SM}.

\section{General solutions}
\label{sec:general-solutions}

As mentioned in Introduction, the Dirac equation (\ref{Dirac-eq-2*2})
for the massless case, $\Delta =0$, and infinite plane was solved
in Refs.~\cite{Lukose2007PRL,Peres2007JPCM}.
Here, we employ a different analytic approach
\cite{Herasymchuk2023PSS} to investigate a finite system.
The main equation (\ref{Dirac-armchair-dimensionless})
describing both zigzag and armchair cases can be rewritten
in the following form
\begin{equation}
\label{chi-tilde}
\partial_\xi \tilde{\chi}_+(\xi)=\left(\tilde{A}+ \tilde{B} \xi \right)\tilde{\chi}_+(\xi),
\end{equation}
where the $2\times2$  $\xi$-independent matrices $\tilde{A}, \tilde{B}$
are, respectively,
\begin{equation}
\label{tilde-A-B}
\tilde{A}  =
\begin{pmatrix}
0 & i(\varepsilon + \delta)\\
i(\varepsilon - \delta) & 0
\end{pmatrix}, \quad
\tilde{B}  =
\begin{pmatrix}
1 & -i\beta \\
-i\beta & -1
\end{pmatrix},
\end{equation}
and the spinor $\tilde{\chi}_+$ is either ${\tilde \psi}_+$
defined below Eq.~(\ref{zigzag-1}) for the zigzag edge case or
${\psi}_+$ defined below Eq.~(\ref{armchair-1})
for the armchair edge case.

While the problem involving the radial electric field \cite{Nimyi2022PRB},
which includes three matrices
${\hat A}/\rho+ {\hat B}+{\hat C} \rho$ with $\rho$ being the
radial variable, cannot  be solved analytically,
the present problem in the crossed uniform  fields in the Cartesian coordinates
is exactly solvable by diagonalizing the matrix $\tilde{B}$, as discussed in detail in \cite{Herasymchuk2023PSS}.
Thus, here we proceed directly to the general solution for
the components of the spinor ${\tilde \chi}_+$,
\begin{equation}
\label{sol-tilde-chi1}
 \hspace*{-0.17cm}
\begin{aligned}
{\tilde \chi}_{+1}(\zeta)
&=i C_{+1}
\left[ \gamma U\left(a-1,\sqrt{2}\zeta \right)-\kappa_{+}
U\left(a,\sqrt{2}\zeta \right)\right]\\
+  i  C_{+2} & \left[ \gamma
V\left(a-1,\sqrt{2}\zeta\right)+\frac{\kappa_{+}}{a-1/2}
V \left(a,\sqrt{2}\zeta \right)\right],
\end{aligned}
\end{equation}

\begin{equation}
\label{sol-tilde-chi2}
 \hspace*{-0.17cm}
\begin{aligned}
{\tilde  \chi}_{+2}(\zeta)
&= C_{+1} \left[  U\left(a-1,\sqrt{2}\zeta \right)-
\gamma \kappa_{+}
U\left(a,\sqrt{2}\zeta \right)\right]\\
+  C_{+2} & \left[  V \left(a-1,\sqrt{2}\zeta \right)+
\frac{\gamma \kappa_{+}}{a-1/2}V \left(a,\sqrt{2}\zeta \right)\right].
\end{aligned}
\end{equation}

Here the solutions are written
in terms of the parabolic cylinder (Weber)
functions $U(a, x)$ and $V(a, x)$, which  depend
on the variable
\begin{equation}
\label{zeta}
\zeta = (1-\beta^2)^{1/4} \xi+ \frac{\beta \varepsilon}{ (1-\beta^2)^{3/4}}
\end{equation}
with $\xi = y/l + kl $ either  $\xi = x/l + kl $ for zigzag or armchair edges.
The integration constants $C_{+1,+2}$ have the restored valley index $+$,
and the following notations are introduced:
\begin{equation}
\label{gamma-kappa}
\gamma=\frac{\beta}{1+\sqrt{1-\beta^2}}, \qquad
\kappa_{\pm}=\frac{\delta \sqrt{1-\beta^2} \pm \varepsilon }
{\sqrt{2}(1-\beta^2)^{3/4}},
\end{equation}
\begin{equation}
\label{a}
a=\frac{1+ 2 \kappa_{-} \kappa_{+}}{2}
=\frac{1}{2}+\frac{\delta^2(1-\beta^2)-\varepsilon^2}{2(1-\beta^2)^{3/2}}.
\end{equation}

The centers of the electron orbital can be defined by the condition
$\zeta =0$, viz,
\begin{equation}
\label{orbital-generic}
y_k = - k l^2 - \frac{\beta l \varepsilon}{1 -\beta^2}.
\end{equation}
The particular relationship between $y_k$ and $\beta$ arises from the dependence of energy $\varepsilon$ on $\beta$. Substituting the spectrum
(\ref{LL-collapse}) \{see also Eq.~(S4) in SM \cite{SM}\}
in the last equation one can see that for an infinite system
\begin{equation}
\label{orbital-infinite}
y_k = - k l^2 - \frac{\beta l \mbox{sgn} \, (\varepsilon_n) \sqrt{2n (1-\beta^2)^{1/2} + \delta^2}}{(1 - \beta^2)^{1/2}}.
\end{equation}
This illustrates that the electron and hole orbits become open
on the opposite sides  as $|\beta| \to 1$.

The derivation of the spectrum (\ref{LL-collapse}) for an infinite system, which utilizes the specific asymptotic behavior of the Weber parabolic cylinder functions, is included in Sec.~S1 within the SM \cite{SM}
for completeness.
We also included there in Secs.~S2 and S3 within the SM \cite{SM}
the results for spectra of the ribbons with the zigzag and armchair edges, respectively.
The  symmetry relations (S11) and (S15) (see the SM \cite{SM})
for the energy spectra of these ribbons are also provided.
Our numerical results support an observation made in Introduction on the base of  Ref.~\cite{Lukose2007PRL} and illustrate that the electron-like levels seem to be denser near $-kl \approx W/l$ edge, while the hole-like are denser for $k l \approx 0$.

To study the level behavior by analytic methods, we simplify the problem by considering the semi-infinite geometry in Sec.~\ref{sec:zigzag-semi}. In particular, this enables us to investigate the specific of the Landau level collapse in the restricted geometry.


\section{Half-plane with the zigzag edge}
\label{sec:zigzag-semi}

On a half-plane, normalizable wave functions can be expressed solely
in terms of the function $U(a, x)$, which, as mentioned above, decays exponentially
as $x \to \infty$. On the other hand, the function $V(a, z)$ grows exponentially in both directions as $x \to \pm \infty$ (see  Refs.~\cite{SM,Herasymchuk2023PSS}).
This enables one to set $C_{+2}=0$ in the solutions
(\ref{sol-tilde-chi1}) and (\ref{sol-tilde-chi2})
the $\mathbf{K}_+$ valley.
However, unlike the case of an infinite plane, on a half-plane, there is no restriction on the parameter $a$ being a negative half-integer.

The zigzag boundary condition (\ref{boundaryZ-W}) at $y=W \to \infty$ is naturally fulfilled as a consequence of the asymptotic behavior of $U(a,x\to \infty)$.
The remaining boundary condition (\ref{boundaryZ-0}) at $y=0$ leads to
the requirement that the term with $C_{+1}$ in Eq.~(\ref{sol-tilde-chi1})
must be zero.  The latter condition results in the following equation
for the spectrum for the $\mathbf{K}_+$ valley:
\begin{equation}
\label{spectrum-zigzag+half}
\gamma U \left(a-1, \sqrt{2} \zeta (0) \right)
- \kappa_{+} U\left(a, \sqrt{2} \zeta (0) \right) =0.
\end{equation}
Here $\zeta (0)  =  (1-\beta^2)^{1/4} k l
+ \beta \varepsilon (1-\beta^2)^{-3/4}$.
Using the asymptotics Eq.~(S1) within the SM \cite{SM} one can
verify that Eq.~(\ref{spectrum-zigzag+half})
also follows from Eq.~(S5) within the SM \cite{SM}
for the spectrum on the ribbon in the limit $W \to \infty$.

Similarly, for the $\mathbf{K}_-$ valley, the  boundary condition (\ref{boundaryZ-0}) at $y=0$ leads to the following equation:
\begin{equation}
\label{spectrum-zigzag-half}
U \left(a-1, \sqrt{2} \zeta (0) \right)
+ \gamma \kappa_{-} U\left(a, \sqrt{2} \zeta (0) \right) =0.
\end{equation}

\begin{figure}[!ht]
\includegraphics[width=.42\textwidth]{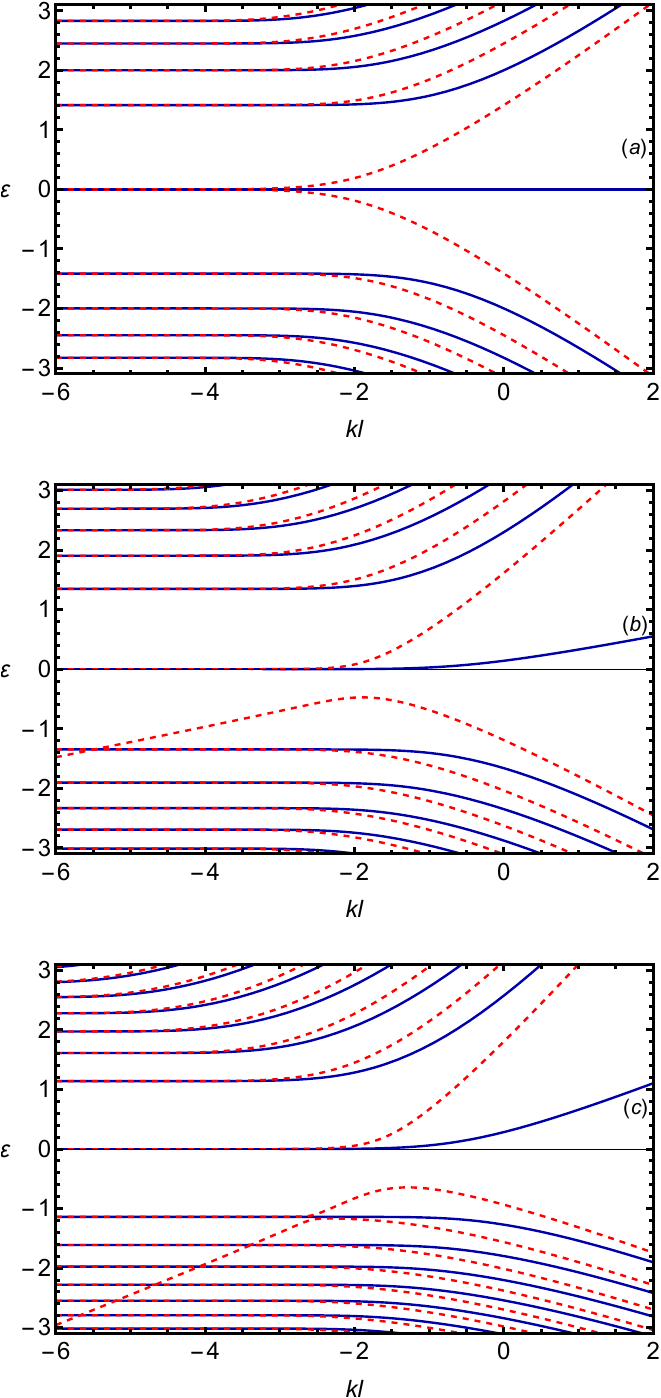}
\caption{The energy spectra $\varepsilon(k)$
of the first few Landau levels near the zigzag edge of a graphene
half-plane for the gapless ($\delta=0$) case. The solutions for the
$\mathbf{K}_+$ and $\mathbf{K}_-$ valleys are shown by the
solid (blue) and dashed (red) lines, respectively.
(a) $\beta =0$; (b) $\beta =0.25$; and (c) $\beta =0.5$
}
\label{fig:zigzag-spectrum}
\end{figure}

\begin{figure}[!ht]
\includegraphics[width=.42\textwidth]{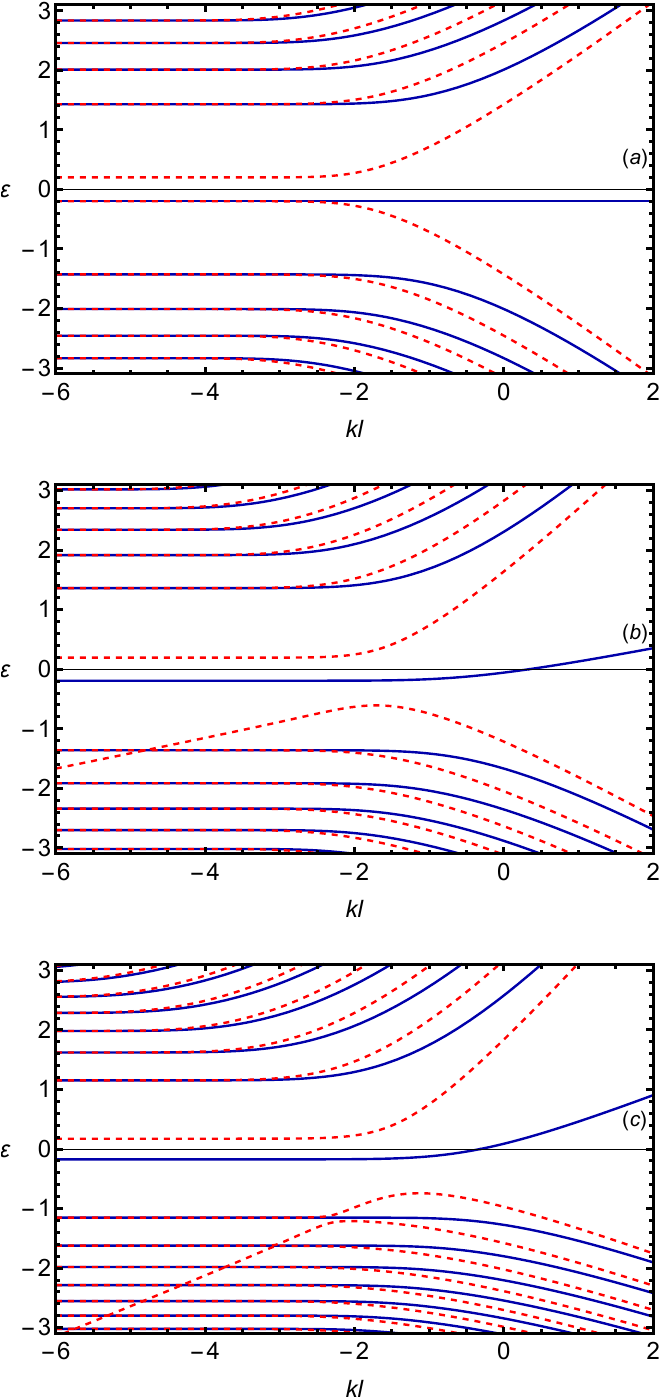}
\caption{The energy spectra $\varepsilon (k)$
of the first few Landau levels near the zigzag edge of  graphene
for the gapped ($\delta=0.2$) case.
The panels (a), (b)  and (c) are for the same
values of $\beta$ as in  Fig.~\ref{fig:zigzag-spectrum}.}
\label{fig:zigzag-gapped-spectrum}
\end{figure}

Equations~(\ref{spectrum-zigzag+half}) and (\ref{spectrum-zigzag-half})
determine dimensionless energies $\varepsilon_\alpha=\varepsilon_n(kl)$
as functions of quantum numbers $\alpha\equiv (n, k)$.
The corresponding spectra
are computed numerically and
presented  for the gapless and gapped cases in
Figs.~\ref{fig:zigzag-spectrum} and \ref{fig:zigzag-gapped-spectrum},
respectively. Notice that, unlike the figures within the
SM \cite{SM}, we present the dimensionless energy
$\varepsilon(k)$ herein. In contrast to the energy $\mathcal{E}$,
this representation excludes the linear in $k$ part.
This enables a clearer presentation of the results when
considering higher values of $\beta$.
In particular, in Fig.\ref{fig:zigzag-spectrum}~(c),
it is observed that the hole-like levels converge more rapidly; specifically,
there are seven visible hole-like levels compared to only six electron-like levels.

In contrast to Figs.~S1 and S2 within the SM  \cite{SM}
where we showed the $\mathbf{K}_{+}$ and $\mathbf{K}_{-}$ valleys on separate
panels, here we superimpose both valleys on the same
panel to allow a direct comparison of the
corresponding energy levels. This is possible, because
in the continuum model  the wave vector $k$ is counted from $K_{\pm x}$
values. The negative values of $k$ correspond to the bulk,
while the edge is at $k=0$.

Since a half-plane geometry is considered, for a finite
$\beta$ the energies $\varepsilon (k)$ tend
to the constant values in the bulk as $kl \to -\infty$.
The presence of the edge at $y=W$ modifies this behavior:
the hole-like levels including the
blue line (the $\mathbf{K}_+$ solution) that goes to zero
for $kl \to -\infty$ would go downward, while the
electron-like levels go upward.
Furthermore, the degeneracy of the solutions for the $\mathbf{K}_+$
and $\mathbf{K}_-$ valleys would be lifted near the other edge
as one can see in Figs.~S1 and S2 (see the SM  \cite{SM}).

The sole dispersive curve for $kl \to -\infty$, given that $\varepsilon (k)$ is considered, corresponds to the dispersionless state observed in the full
spectrum $l \mathcal{E} (k)/(\hbar v_F)$. Specifically, this is the zero-energy lower branch of
the $\mathbf{K}_-$ valley spectrum in the ribbon geometry.
It describes the surface states mentioned in Sec.~S2~A (see the SM \cite{SM}).
The second edge of the ribbon supports a second dispersionless mode
which is absent in a half-plane geometry.
The surface state in the semi-infinite geometry is briefly discussed in
Sec.~\ref{sec:dispersionless}.
The analytical consideration of the Landau level spectrum in zero
electric field was presented in \cite{Herasymchuk2023PSS} (see also \cite{Romanosvsky2011PRB}). Since for $E=0$ the presented below expressions reduce
to the corresponding formulas from Ref.~\cite{Herasymchuk2023PSS}, we
include them in Sec.~S4 within the SM \cite{SM} for a reference.

\subsection{Landau levels in the bulk for $\beta \neq 0$}

It is possible to obtain the approximate
analytic expressions that generalize
Eqs.~(S17) and  (S18) from the SM \cite{SM}
in the bulk, specifically for $y_0 = - k l^2 \gg l$ or
$k l \to - \infty$. The
$\mathbf{K}_+$ valley solutions are
\begin{equation}
\label{sol0-K+beta}
\varepsilon_{+,0}=-\delta \left(1-\beta^2\right)^{\frac{1}{2}}+\frac{\gamma \left(1-\beta^2\right)^{\frac{3}{4}}}{\sqrt{\pi}} e^{- \left(1-\beta^2\right)^{\frac{1}{2}} (kl)^2}
\end{equation}
and
\begin{equation}
\label{sol-K+beta}
\begin{split}
& \varepsilon_{+,n}^2-\delta^2 \left(1-\beta^2\right)=2n\left(1-\beta^2 \right)^{\frac{3}{2}}
\left[1+\frac{2^n (kl)^{2n}}{\sqrt{\pi} n!} \right. \\
& \times \left.
\frac{\gamma \left(1-\beta^2\right)^{\frac{2n+1}{4}}}{\sqrt{2n\left(1-\beta^2\right)^{\frac{1}{2}}+\delta^2}+\delta} e^{-\left(1-\beta^2\right)^{\frac{1}{2}} (kl)^2} \right]
\end{split}
\end{equation}
with  $n =1,2, \ldots$. The corresponding solutions for
$\mathbf{K}_-$ valley are
\begin{equation}
\label{sol0-K-beta}
\varepsilon_{-,0}=\delta \left(1-\beta^2\right)^{\frac{1}{2}}+\frac{\left(1-\beta^2\right)^{\frac{3}{4}}}{\sqrt{\pi}\gamma} e^{- \left(1-\beta^2\right)^{\frac{1}{2}} (kl)^2}
\end{equation}
and
\begin{equation}
\label{sol-K-beta}
\begin{split}
& \varepsilon_{-,n}^2-\delta^2 \left(1-\beta^2\right)=2n\left(1-\beta^2 \right)^{\frac{3}{2}} \left[1+\frac{2^n (kl)^{2n}}{\sqrt{\pi} \gamma n!} \right. \\
\times
& \left. \frac{ \left(1-\beta^2\right)^{\frac{2n+1}{4}}}{\sqrt{2n\left(1-\beta^2\right)^{\frac{1}{2}}+\delta^2}-\delta} e^{-\left(1-\beta^2\right)^{\frac{1}{2}} (kl)^2} \right]
\end{split}
\end{equation}
with  $n =1,2, \ldots$.

Note that since $\beta \neq 0$ in the derivation of the asymptotic
(\ref{sol0-K+beta}), (\ref{sol-K+beta}), (\ref{sol0-K-beta})
and  (\ref{sol-K-beta}), we have taken into account
the contribution of another parabolic cylinder function in
Eqs.~(\ref{spectrum-zigzag+half})
and (\ref{spectrum-zigzag-half}) as compared to the solution
of Eqs.~(S16) from the SM \cite{SM}.
Thus, Eqs.~(\ref{sol-K+beta}) and (\ref{sol0-K+beta}) are valid for any value of
$\beta \in (-1,1)$.

On the contrary, the contribution of $\gamma \kappa_- U(a,\sqrt{2} \zeta(0))$ in
Eq.~(\ref{spectrum-zigzag-half}) for the
$\mathbf{K}_{-}$ valley spectrum leads to cancellation of a correction of the order $\sim  \left(y_0/l\right)^{2n+1}e^{-\left(1-\beta^2\right)^{\frac{1}{2}} y_0^2/l^2}$, and hence, the correction of the lower order survives in Eq.~(\ref{sol-K-beta}), $\sim \gamma^{-1} \left(y_0/l\right)^{2n}e^{-\left(1-\beta^2\right)^{\frac{1}{2}} y_0^2/l^2}$.
Therefore, in the case of the $\mathbf{K}_{-}$ valley, the limit $\beta \rightarrow 0$ in Eq.~(\ref{sol-K-beta}) [and Eq.~(\ref{sol0-K-beta})]  is not applicable, hence it is
valid for values of $\beta \in (-1,-\epsilon] \cup [\epsilon,1)$ with $\epsilon>0$.

As can be seen from the Eqs.~(\ref{sol-K+beta}), (\ref{sol0-K-beta})
and  (\ref{sol-K-beta}), the asymmetry of the electron-like and hole-like Landau levels disappears
in the limit $kl \rightarrow -\infty$. However, closer to the edge ($kl=0$), the difference between
the energy values $\varepsilon(k)$ tends to $0$ as $\beta$ approaches $1$ for the hole-like levels.
For the electron-like levels, the difference tends to a constant value as demonstrated in
Figs.~\ref{fig:zigzag-spectrum} and \ref{fig:zigzag-gapped-spectrum},
it is also shown explicitly in Fig.~\ref{fig:collapse-edge}  for the energy spectra at $kl=0$.

\subsection{Landau levels and probability density for $\beta \neq 0$}
\label{sec:zigzag-finiteE-edge}

\begin{figure}[!ht]
\includegraphics[width=.42\textwidth]{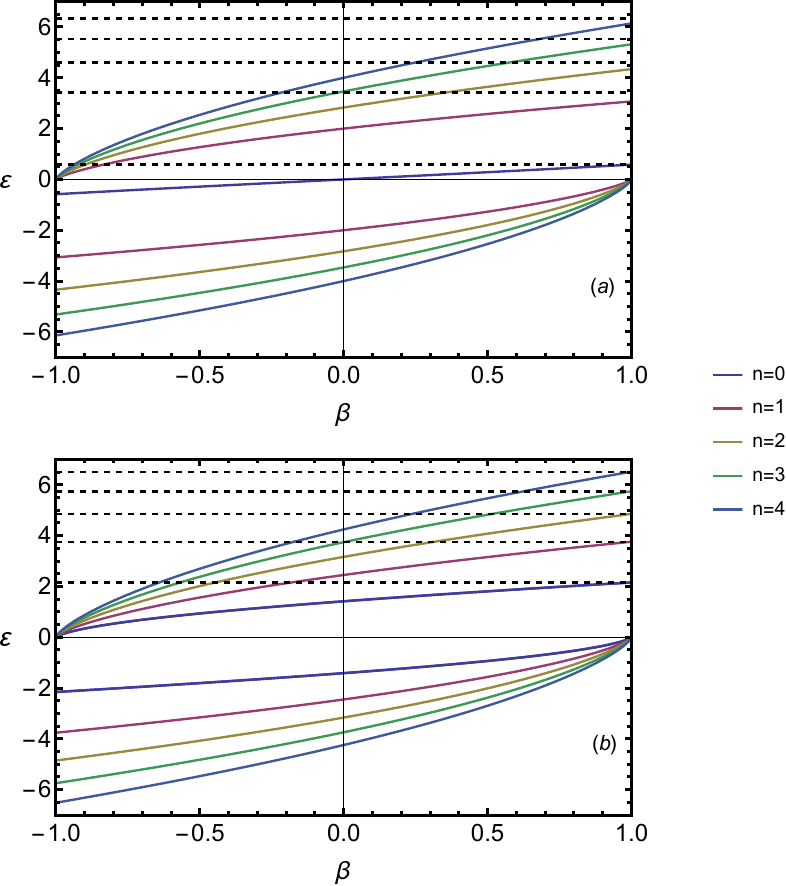}
\caption{The energy spectra $\varepsilon(kl =0)$ at the zigzag edge vs
electric field in terms of $\beta = c E / (v_F H)$ for $\delta =0$ for
first few Landau levels. The dashed lines correspond to the energies
given by Eq.~(\ref{e-levels-set-beta=1}):
(a) For $\mathbf{K}_+$ valley, (b) For $\mathbf{K}_-$
valley. }
\label{fig:collapse-edge}
\end{figure}

We begin by presenting in Fig.~\ref{fig:collapse-edge} numerical solutions
of Eqs.~(\ref{spectrum-zigzag+half}) and (\ref{spectrum-zigzag-half})
for the energy spectra at the zigzag edge, $\varepsilon(kl =0)$,
and $\delta =0$.
These solutions are depicted as functions of $-1 \leq \beta \leq 1$ for the
first few Landau levels.
For $\beta =0$ the values of the energies agree
with the lower lines of Eqs.~(S17) and (S18) from the SM \cite{SM}.
One can see that for the $\mathbf{K}_+$
valley the $n=0$ level does not collapse,
while for the $n \neq 0$ levels the electron-like solutions with $\varepsilon >0$
merge to one point and collapse only when $\beta \to -1$.  For $\beta \to 1$ the
level energies tend to the different values and thus the electron-like levels
do not collapse at the edge.
On the contrary, the hole-like solutions with $\varepsilon < 0$
merge and collapse for $\beta \to 1$, while for $\beta \to -1$ these levels
do not collapse.
For the $\mathbf{K}_-$ valley all $n \geq 0$ levels behave similarly to
$n \neq 0$ levels in the $\mathbf{K}_+$ valley.
We verified that spectra exhibit the  same behavior  for finite values of $\delta$.

The observed behavior can be explained qualitatively using Eq.~(\ref{orbital-generic})
for the centers of the electron orbital. Indeed, the Landau level
collapse occurs when the corresponding center of electron or hole orbit
remains in the bulk ($y >0$), while $y_k \to \infty $.
The behavior of the levels on the opposite edge is interchanged
when compared to the edge considered in this section, because
in this case the bulk is for $y <0$.

A further understanding of electron's behavior can be reached by directly addressing
the wave functions and the corresponding probability density
\begin{equation}
\label{probability}
\rho_{\eta,n}^{(\alpha)}\left( y\right) = \Psi_{\eta,n}^{(\alpha)\dagger}(\mathbf{r},k)
\Psi_{\eta,n}^{(\alpha)}(\mathbf{r},k).
\end{equation}
where $\Psi_{\eta,n}^{(\alpha)}(\mathbf{r},k)$ is the normalized wave function of
the states, characterized by the discrete quantum numbers $(\eta, n, \alpha)$ and
wave vector $k$, explicitly written in Appendix~\ref{sec:Appendix-probability}.
Here $n$ is  the Landau level index and $\alpha  =e,h$ is the electon-like ($\varepsilon >0$)
or hole-like ($\varepsilon <0$) levels.

\begin{figure}
\includegraphics[width=.42\textwidth]{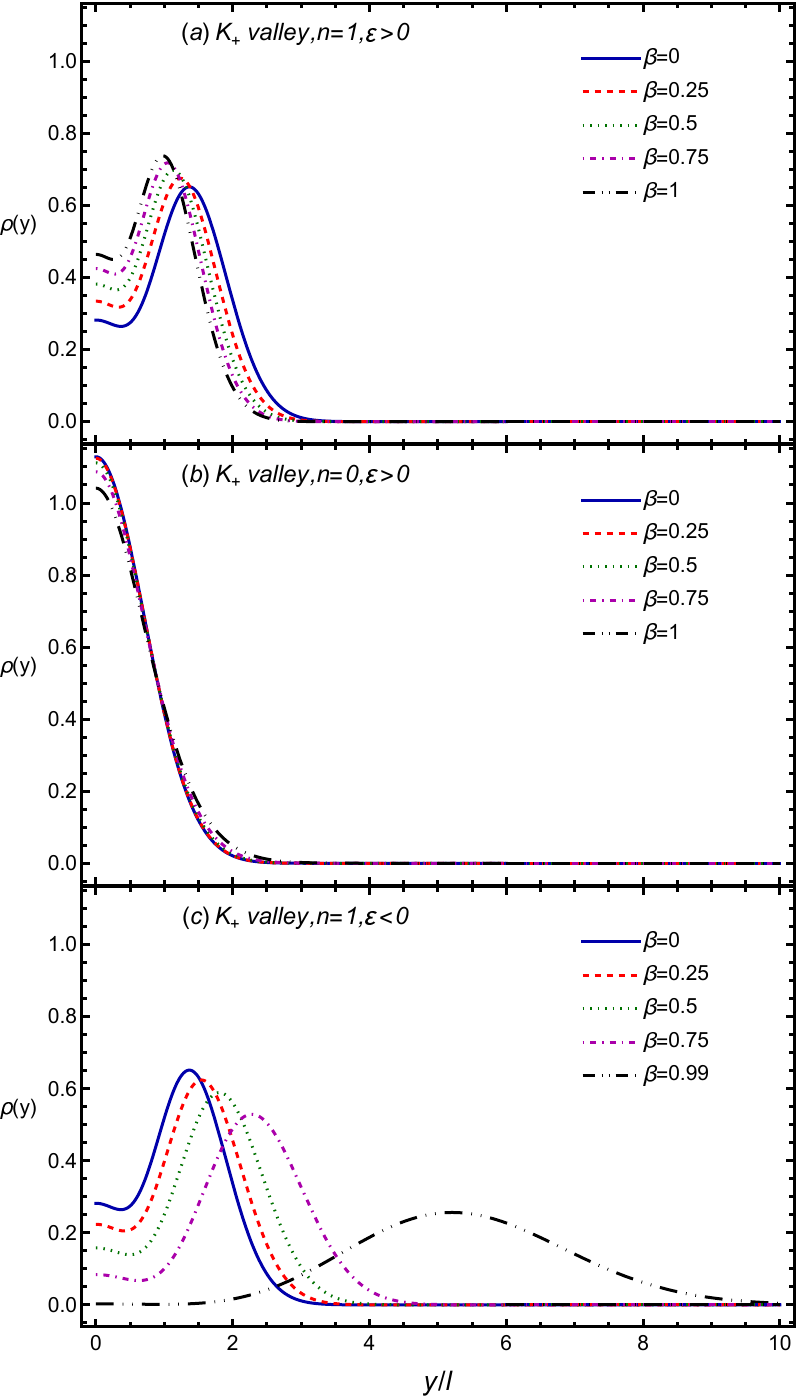}
\caption{The probability densities for $\mathbf{K}_{+}$ valley, $kl=0$, $\delta =0$,
and five different values of $\beta = 0, 0.25, 0.5, 0.75, 1$ (see the remark concerning $\beta =1$ in the main text). (a) $n=1,\,\,\varepsilon>0$; (b) $n=0,\,\,\varepsilon>0$; and (c) $n=1,\,\,\varepsilon<0$. }
\label{fig:probability-K+}
\end{figure}

\begin{figure}
\includegraphics[width=.42\textwidth]{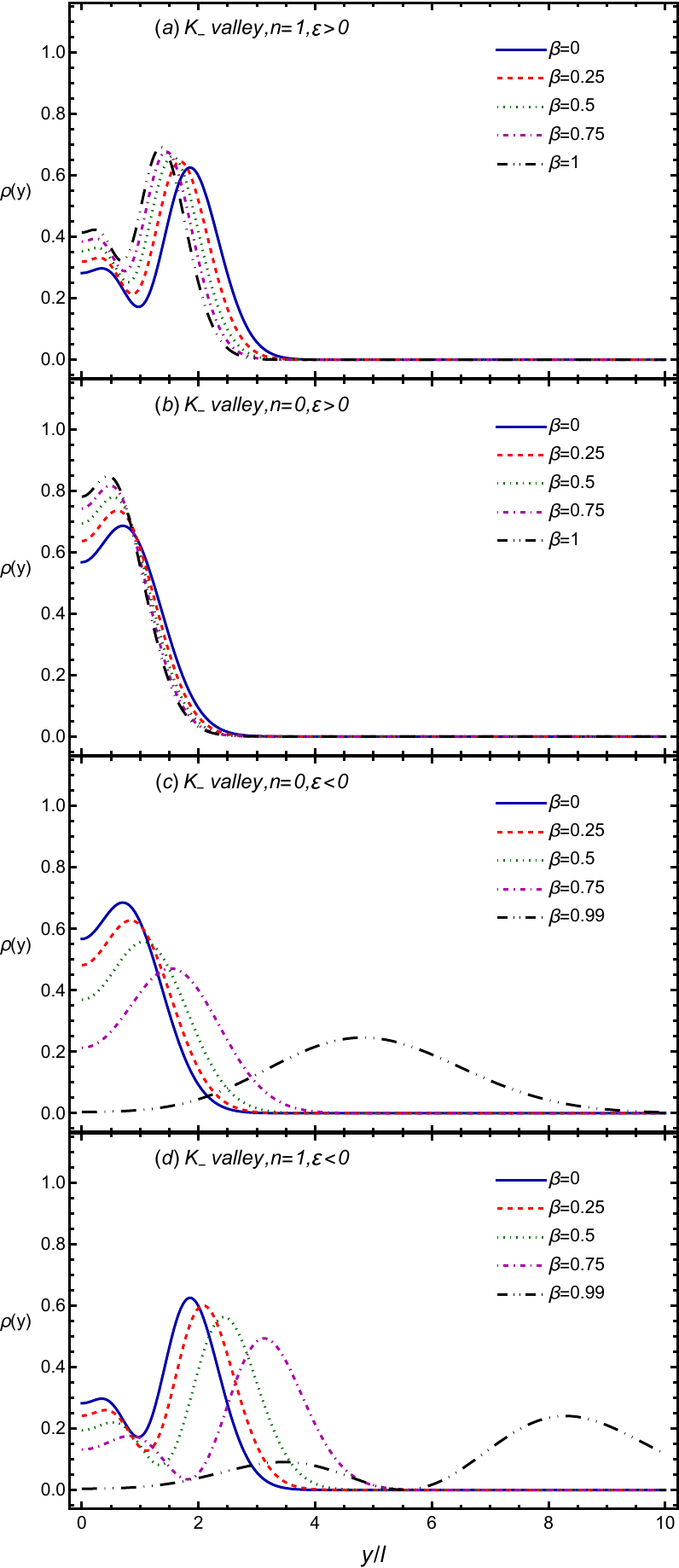}
\caption{The probability densities for $\mathbf{K}_{-}$ valley, $kl=0$, $\delta =0$, and
five different values of $\beta = 0, 0.25, 0.5, 0.75, 1$ (see the remark concerning $\beta =1$ in the main text).
(a) $n=1,\,\,\varepsilon>0$; (b) $n=0,\,\,\varepsilon>0$; (c) $n=0,\,\,\varepsilon<0$; and (d) $n=1,\,\,\varepsilon<0$.
}
\label{fig:probability-K-}
\end{figure}

\begin{figure}
\includegraphics[width=.42\textwidth]{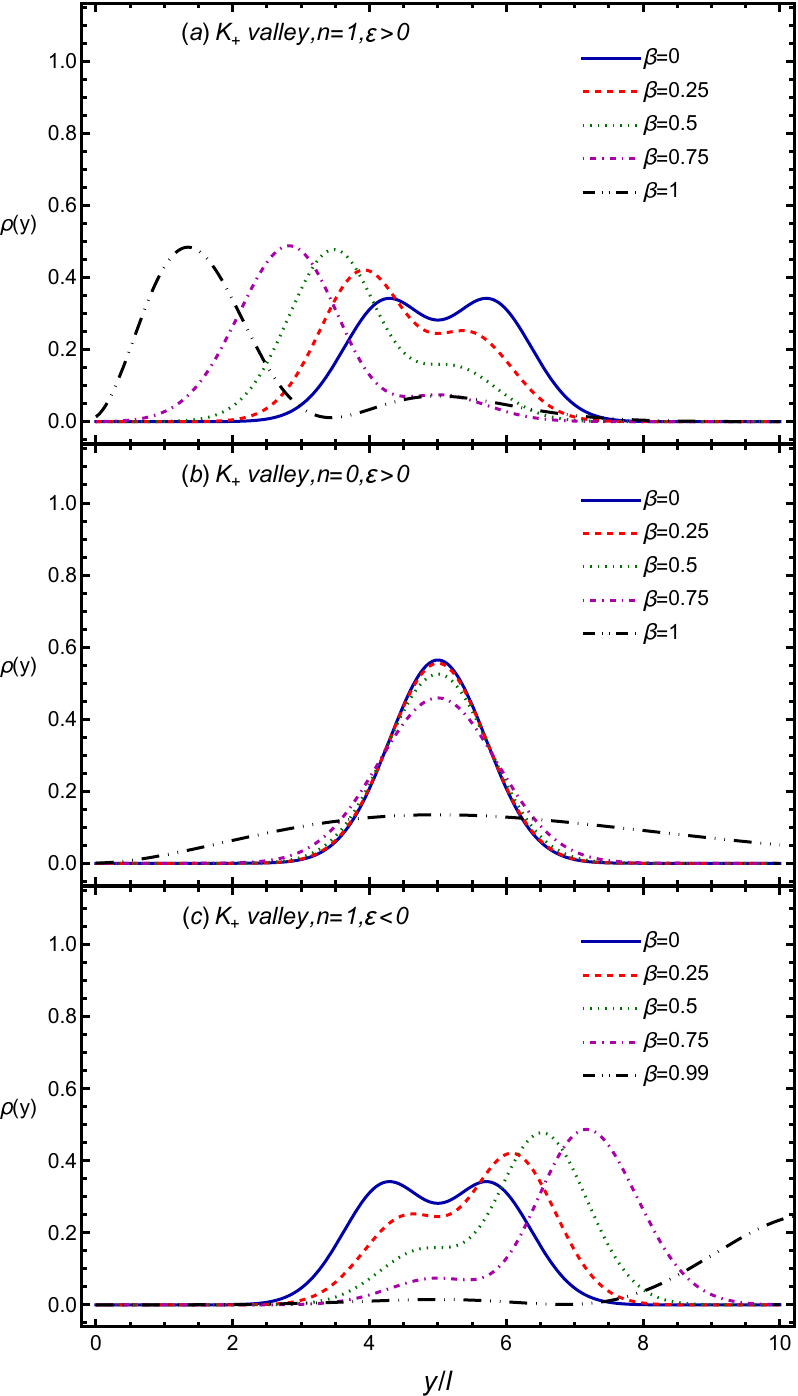}
\caption{The probability densities for $\mathbf{K}_{+}$ valley,  $kl=-5$, $\delta =10^{-7}$,
and five different values of $\beta$. (a) $n=1,\,\,\varepsilon>0$; (b) $n=0,\,\,\varepsilon>0$; and (c) $n=1,\,\,\varepsilon<0$. }
\label{fig:probability-K+bulk}
\end{figure}

\begin{figure}
\includegraphics[width=.42\textwidth]{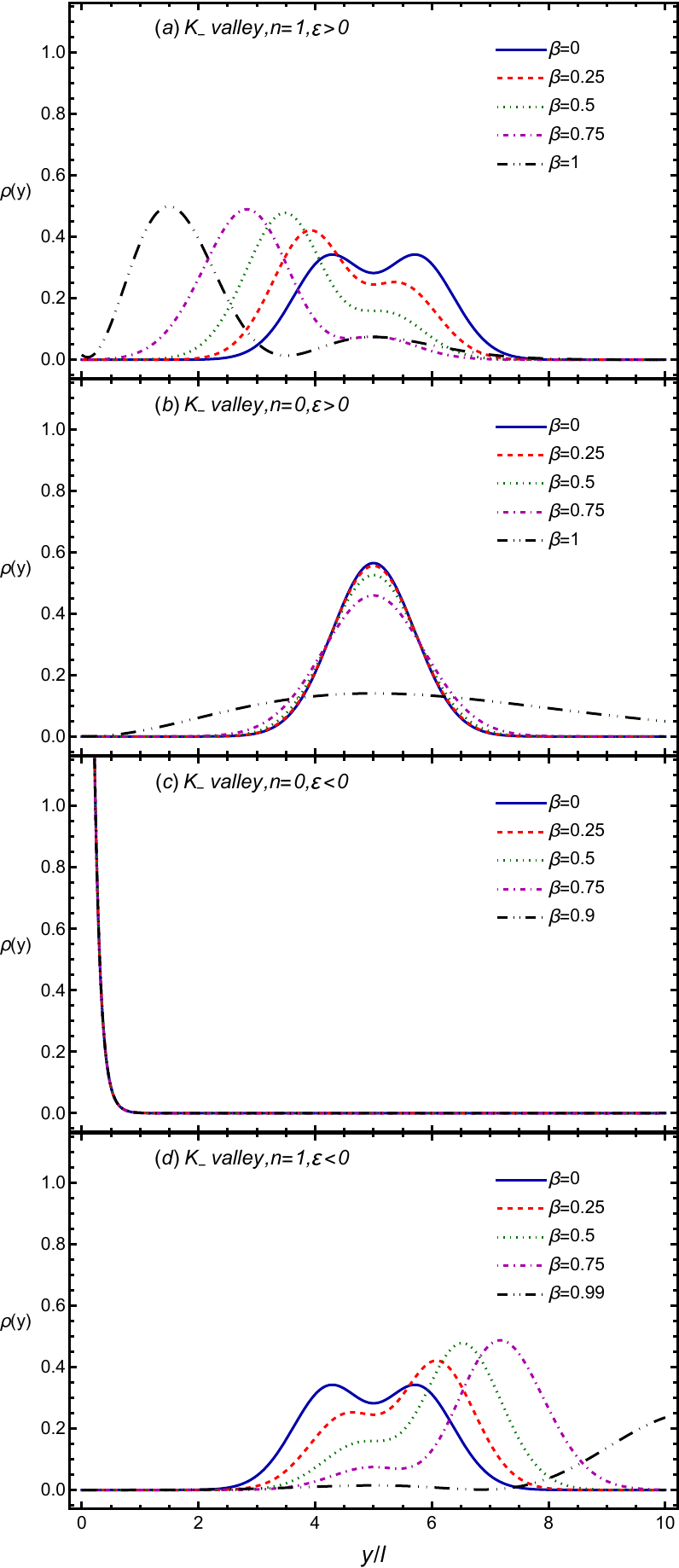}
\caption{The probability densities for $\mathbf{K}_{-}$ valley, $kl=-5$, $\delta =10^{-7}$,  and
five different values of $\beta$.
(a) $n=1,\,\,\varepsilon>0$; (b) $n=0,\,\,\varepsilon>0$; (c) $n=0,\,\,\varepsilon<0$; and (d) $n=1,\,\,\varepsilon<0$.
}
\label{fig:probability-K-bulk}
\end{figure}

The corresponding probability density distributions for the states close to  zero energy
as a function of a distance $y$ from the half-plane edge are plotted
in Figs.~\ref{fig:probability-K+}--\ref{fig:probability-K-bulk} for $\mathbf{K}_{\pm}$ valleys:
$kl=0$ (close to the edge) in Figs.~\ref{fig:probability-K+} and \ref{fig:probability-K-}, and $kl =-5$ (in the bulk)
in Figs.~\ref{fig:probability-K+bulk} and \ref{fig:probability-K-bulk}, respectively.
Note that according to the discussion presented in Sec.~\ref{sec:half-critical},
for the $\varepsilon<0$ case, instead of $\beta=1$ we took a smaller value
$\beta=0.99$ to have normalizable solutions.
The results shown support the above statement for the orbit centers.
Indeed,  the maximal probability density of the non-collapsing electron-like levels
remain near the edge, while the wave functions of the collapsing hole-like levels move into the bulk
as the value of $\beta,\,\,0\le\beta\le1$ increases.
To remove the degeneracy of the lowest $n=0$ Landau level in Figs.~\ref{fig:probability-K+bulk} and
\ref{fig:probability-K-bulk}, the gap
$\delta =10^{-7}$ is taken.

The number of panels for the $\mathbf{K}_{+}$ point in Figs.~\ref{fig:probability-K+} and
\ref{fig:probability-K+bulk} is smaller than for $\mathbf{K}_{-}$ point in
Figs.~\ref{fig:probability-K-} and \ref{fig:probability-K-bulk} reflecting the fact that
the $n=0$ level in
Figs.~\ref{fig:zigzag-spectrum} and \ref{fig:zigzag-gapped-spectrum} has
only one branch in the spectrum for $\mathbf{K}_{+}$ valley and two branches for $\mathbf{K}_{-}$ valley,
respectively.
For the $n=1$ level, the motion of the maxima of the probability density with the increase
of positive $\beta$ is similar in both cases, $kl=0$ and $kl=-5$. For the $n=0$
level, the bulk-like electron states ($\varepsilon >0$) at $\mathbf{K}_{\pm}$ valleys eventually widen,
while the dispersionless edge state with  $\varepsilon <0$ [Fig.~\ref{fig:probability-K-bulk}~(c)]
does not move (see further discussion in Sec.~\ref{sec:dispersionless}).

\subsubsection{Landau levels at the edges for $\beta \neq 0$}

Now we proceed to the discussion of the analytical results obtained for
Eqs.~(\ref{spectrum-zigzag+half}) and (\ref{spectrum-zigzag-half}) in the limit $k l \to 0$.
Finding a straightforward generalization of solutions
(S17) and (S18) from the SM \cite{SM}
near the edges, specifically when $y_0 \ll l$ or $k l \to 0$, proves to be
challenging.
Nevertheless, it is possible to consider analytically two limiting cases: $|\beta| \ll 1$ and $|\beta| \rightarrow 1$. The details of the derivation
are described in Appendix \ref{sec:Appendix-zigzag}, where we obtain that
for $|\beta| \ll 1$ the solutions for the $\mathbf{K}_{\pm}$ valleys
read, respectively,
\begin{equation}
\label{sol-K+beta-small}
\begin{aligned}
\varepsilon_{+,0} &=- \delta + \frac{1}{\sqrt{\pi}} \beta +o(\beta)  \\
\varepsilon_{+,n} &=\pm \sqrt{4 n+\delta^2}  \left[1+ \frac{\Gamma\left(n+\frac{1}{2}\right)}{\Gamma\left(n\right)}
\frac{2  }{\pi
\left( 4n+\delta^2 \right)} \right. \\
& \times \left.  \left( \frac{ 1 }{  \delta \pm \sqrt{4n+\delta^2} }  \pm 2  \sqrt{4n+\delta^2}\right) \beta +o(\beta) \right]
\end{aligned}
\end{equation}
with  $n =1,2, \ldots$, and
\begin{equation}
\label{sol-K-beta-small}
\begin{split}
 \varepsilon_{-,n} & =  \pm \sqrt{2 (2n+1)+\delta^2}  \left[ 1- \frac{\Gamma\left(n+\frac{3}{2}\right)}{\Gamma\left(n+1\right)} \right.\\
& \times \frac{ 2 }{\pi \left( 2 (2n+1)+\delta^2 \right)}  \left(
\frac{ 1 }{  \delta \pm \sqrt{2(2n+1)+\delta^2}} \right.   \\
& \left.  \mp 2\sqrt{2 (2n+1)+\delta^2} \Biggr) \beta +o(\beta) \right]
\end{split}
\end{equation}
with  $n =0,1, \ldots$.

The details of the derivation for the limit $|\beta| \rightarrow 1$
are also described in Appendix \ref{sec:Appendix-zigzag}.
For the sake of simplicity we consider only the gapless case, $\delta=0$,
when Eqs.~(\ref{spectrum-zigzag+half}) and (\ref{spectrum-zigzag-half})
possess the following symmetry: $\varepsilon(\beta) = -\varepsilon (-\beta)$.
Then it is convenient to consider only the electron-like levels,
while  the properties of the hole-like levels follow from
the symmetry. We found that for $\beta \to 1$
the electron-like levels {\em do not collapse} at
the edge, $kl =0$, and their energies  tend
to the following values
\begin{equation}
\label{e-levels-set-beta=1}
\varepsilon_{\pm,n} = \sqrt{2 \varkappa_{\pm,n}^3} , \qquad  n =0,1 \ldots.
\end{equation}
Here $\varkappa_{\pm,n}$ are the roots of the equations $f_{\pm}(x) =0$,
where the functions
\begin{equation}
\label{f-pm}
f_\pm(x)=\text{Ai}' \left( -x^2  \right) \pm x \text{Ai}\left( -x^2  \right)
\end{equation}
are expressed in terms of the Airy function
$\text{Ai}(z)$ and its derivative.
One can see that for  $x > 0$ they have  infinite sets of zeros
$\varkappa_{\pm,n}$, $n=0,1,2,\ldots$
The energies given by Eq.~(\ref{e-levels-set-beta=1})
for the $\mathbf{K}_{\pm}$ valleys are shown
in Figs.~\ref{fig:collapse-edge}~(a) and (b), respectively, confirming the
consistency between calculations done for the original equations
for the spectra and their $\beta \to 1$ limit.

For large $n$ one can use the expansions
(\ref{Airy-expand}) reducing equations for the spectrum
to the trigonometric ones [see Appendix~\ref{sec:Appendix-zigzag},
Eqs.~(\ref{spectrum-trig+}) and (\ref{spectrum-trig-})]
and obtain the following approximate expressions:
\begin{equation}
\label{e-levels-set-beta=1-approx}
\begin{cases}
\varepsilon_{+,n}^{\mathrm{app}} = \sqrt{3 \pi n} , &  n =1,2 \ldots, \\
\varepsilon_{-,n}^{\mathrm{app}} = \sqrt{3 \pi \left(n+\frac{1}{2} \right)} , & n =0,1, \ldots \\
\end{cases}
\end{equation}
As discussed in Appendix~\ref{sec:Appendix-zigzag} there is a rather good
agreement with the results obtained solving numerically
the full equation involving Airy function.

On the other hand, we obtained (see Appendix~\ref{sec:Appendix-zigzag})
that in the limit $\beta \to -1$
the electron-like levels {\em collapse} at
the edge, $kl =0$, and their energies are the
\begin{equation}
\label{e-levels-set-beta=-1}
\begin{cases}
\varepsilon_{+,n} =  \nu_{2n}   (1-\beta^2)^{3/4} , &  n =1,2 \ldots, \\
\varepsilon_{-,n} = \nu_{2n+1}   (1-\beta^2)^{3/4} , & n =0,1, \ldots, \\
\end{cases}
\end{equation}
where $\nu_{2n}$ and $\nu_{2n+1}$ are the roots of the equation
$f(\nu) =0$ with the function $f(\nu)$ given by Eq.~(\ref{f-nu}).
Note that the collapse point with $\varepsilon =0$ and $\beta =-1$
does not belong to the spectra, because the original equations in this case
do not have a nontrivial solution (see Sec.~\ref{sec:critical}).
The lowest, $n=0$, Landau level for the $\mathbf{K}_{+}$ valley should be considered separately, since it does not collapse at the edge for any value of $\beta$.
We obtain for $\delta=0$ the following approximate expression:
\begin{equation}
\varepsilon_{+,0}=  \frac{1}{\sqrt{\pi}} \beta+o(\beta).
\end{equation}

One may observe an analogy with the specifics of the
Landau level collapse in a  field configuration involving a constant in-plane
radial electric field \cite{Nimyi2022PRB}. Although an
infinite system, rather than a restricted geometry, was considered
in \cite{Nimyi2022PRB}, the electron- and hole-like
Landau levels collapse differently depending
on the direction of the electric field and the
angular momentum quantum number.

Finally, we note that the analysis done for the edge, $kl =0$,
can also be extended for finite values $kl$.

\subsection{Surface mode in a finite electric field}
\label{sec:dispersionless}

The dispersionless surface states are localized at the boundaries \cite{Fujita1996JPSJ,Brey2006aPRB,Brey2006bPRB,Ababnin2006PRL}, and,
along with the $n=0$ Landau level, constitute the degenerate states.
The electric field lifts the degeneracy of
the $n=0$ Landau level and the dispersionless state.
Indeed, in Figs.~\ref{fig:zigzag-spectrum}~(b) and (c) one observes
the splitting of the two red (dashed) curves for the $\mathbf{K}_-$ valley that
merge to zero energy in Fig.~\ref{fig:zigzag-spectrum}~(a)
as $kl \to -\infty$. The upper curve corresponds to the
dispersing $n=0$ level, while the lower curve
is related to the dispersionless surface state.
Indeed, as one can see in Fig.~\ref{fig:probability-K-bulk}~(b)
for $\varepsilon >0$ the probability density characterizes the
bulk-like state. On the other hand, the probability density for $\varepsilon <0$
in Fig.~\ref{fig:probability-K-bulk}~(c) is localized at the edge
and corresponds to the dispersionless state. It does not change with the increase of $\beta$,
while the bulk-like state widens. (See also Ref.~\cite{Romanosvsky2011PRB},
where the separate components of the spinors are plotted.)
As one can see, in the case of the ribbon \{see Figs.~S1~(c) and S1~(d) within the SM \cite{SM}\}
in the $\mathbf{K}_+$ valley this state evolves
into the dispersing lowest $n=0$  Landau level whose energy
decreases as $kl \to -\infty$. In a half-plane geometry the
corresponding blue curve increases linearly as $kl \to -\infty$.

Comparing Figs.~S1~(c) and S1~(d) and
\ref{fig:zigzag-spectrum}~(b) and \ref{fig:zigzag-spectrum}~(c) ,
it is easy to see  the dispersionless mode emerges from a delicate
cancellation with the $\varepsilon$ term that should
show the same dependence on $kl$.
(Recall that the total energy $l \mathcal{E}/(\hbar v_F) = \varepsilon - \beta k l$ and that in Figs.~\ref{fig:zigzag-spectrum} and \ref{fig:zigzag-gapped-spectrum} the value
$\varepsilon $ is plotted, while in \cite{Herasymchuk2023PSS} the value $\mathcal{E}$
is presented in the corresponding figures, see also the discussion in the beginning of
Sec.~\ref{sec:zigzag-semi}.)
The cancelation of these terms even in the presence of an electric field
was proven analytically in \cite{Herasymchuk2023PSS}
by employing the Darwin's expansion of the parabolic cylinder functions
of large order and argument \cite{Abramowitz.book}.

\section{Zigzag ribbon and half-plane in the critical regime, $\beta = \pm 1$}
\label{sec:critical}

At the end of Sec.~\ref{sec:zigzag-finiteE-edge} (see also Appendix~\ref{sec:Appendix-zigzag}) we analyzed the spectrum
in the limit $\beta \to \pm1$ employing the asymptotic
expansion of the parabolic cylinder function
in terms of Airy functions.
One can also derive rather simple equations
for the spectrum for all values of $kl$ in the critical regime
by examining the system of equations (\ref{chi-tilde}) directly
for the $\beta = \pm 1$ case. Then the matrix  $\tilde{A}$ is
given by Eq.~(\ref{tilde-A-B}) and the matrix  $\tilde{B}$
reduces to
\begin{equation}
\tilde{B} =\begin{pmatrix}
 1 & -i \, \text{sgn} \left( \beta \right)\\
-i\, \text{sgn} \left( \beta \right)& -1
 \end{pmatrix}.
\end{equation}
As mentioned in Sec.~\ref{sec:general-solutions}
the system is solved by making
the transformation $\tilde{\chi}_{+} (\xi)=P \chi (\xi)$,
which diagonalizes the matrix  $\tilde{B}$.
For $|\beta| = 1$ the matrix $\tilde{B}$ becomes non-diagnosable, but
choosing the matrix
\begin{equation}
P =\begin{pmatrix}
 1 & i\\
i& 1
 \end{pmatrix}
\end{equation}
we obtain from Eq.~(\ref{chi-tilde}) the system
\begin{equation}
\partial_{\xi} \chi (\xi)=\left( A +B \xi\right) \chi(\xi)
\end{equation}
where
\begin{equation}
A=P^{-1}\tilde{A} P=\begin{pmatrix}
 -\delta & i \varepsilon \\
 i \varepsilon & \delta
 \end{pmatrix},
\end{equation}
and $B \big|_{\beta=\pm}=P^{-1}\tilde{B}\big|_{\beta=\pm}P$ with
\begin{equation}
B \big|_{\beta=+1} = \begin{pmatrix}
 0 & 0\\
-2i & 0
 \end{pmatrix},
\quad
B \big|_{\beta=-1}= \begin{pmatrix}
 0 & 2i\\
0 & 0
 \end{pmatrix}.
\end{equation}
Hence, we obtain the following systems of equations
\begin{equation}
\label{system-critical}
\begin{cases}
\partial_{\xi} \chi_1 =i \varepsilon \chi_2 -\delta \chi_1, \\
\partial_{\xi} \chi_2 =i \varepsilon \chi_1 +\delta \chi_2 -2i \xi \chi_1
\end{cases}
\quad \mbox{for} \quad \beta = +1, \\
\end{equation}
\begin{equation}
\begin{cases}
\partial_{\xi} \chi_1 =i \varepsilon \chi_2 -\delta \chi_1 +2i \xi \chi_2, \\
\partial_{\xi} \chi_2 =i \varepsilon \chi_1 +\delta \chi_2
\end{cases}
\quad \mbox{for} \quad \beta = -1.
\end{equation}
Expressing the component $\chi_2$ via $\chi_1$ for $\beta = +1$
and $\chi_1$ via $\chi_2$ for $\beta = -1$
we arrive at the following systems of equations:
\begin{equation}
\label{system-critical-1}
\begin{cases}
\chi''_1(\xi)=\left[2 \varepsilon \xi -\left( \varepsilon^2 -\delta^2\right) \right] \chi_1(\xi),\\
\chi_2 (\xi)=\frac{1}{i \varepsilon} \left( \partial_{\xi} \chi_1(\xi) +\delta \chi_1(\xi)  \right)
\end{cases}
\quad \mbox{for} \quad \beta = +1, \\
\end{equation}
\begin{equation}
\begin{cases}
\chi_1(\xi) =\frac{1}{i \varepsilon} \left(\partial_{\xi} \chi_2(\xi)-\delta \chi_2 (\xi)\right) \\
\chi''_2(\xi)=\left[-2 \varepsilon \xi -\left( \varepsilon^2 -\delta^2\right) \right] \chi_2(\xi)
\end{cases}
\quad \mbox{for} \quad \beta = -1.
\end{equation}
It is easy to see that for $\beta =\pm 1$ the solutions for $\chi_{1,2}$
can be written in terms of Airy functions.
\begin{widetext}
Thus, for example,
for $\beta =+1$ one has
\begin{equation}
\label{solutions-Airy}
\begin{aligned}
\chi_1(\xi)&=C_{+1} \text{Ai}\left(\frac{\delta ^2-\varepsilon ^2+2 \varepsilon \xi  }{(4\varepsilon^2) ^{1/3}}\right)+C_{+2} \text{Bi}\left(\frac{\delta ^2-\varepsilon ^2+2 \varepsilon \xi  }{(4\varepsilon^2) ^{1/3}}\right)\\
\chi_2(\xi)&=- C_{+1} \frac{i}{\varepsilon }  \left[   \delta \,  \text{Ai}\left(\frac{\delta ^2-\varepsilon ^2+2 \varepsilon \xi  }{(4\varepsilon^2) ^{1/3}}\right)+\frac{2\varepsilon}{(4\varepsilon^2) ^{1/3}} \text{Ai}' \left(\frac{\delta ^2-\varepsilon ^2+2 \varepsilon \xi  }{(4\varepsilon^2) ^{1/3} }\right) \right] \\
&-C_{+2} \frac{i}{\varepsilon }  \left[ \delta \,  \text{Bi}\left(\frac{\delta ^2-\varepsilon ^2+2 \varepsilon \xi  }{(4\varepsilon^2) ^{1/3} }\right)+ \frac{2\varepsilon}{(4\varepsilon^2) ^{1/3}} \text{Bi}'\left(\frac{\delta ^2-\varepsilon ^2+2 \varepsilon \xi  }{(4\varepsilon^2) ^{1/3}}\right) \right].
\end{aligned}
\end{equation}
\end{widetext}
Similarly, one obtains the solution for $\beta = -1$.
Then one can derive the expressions for the spinors
$\tilde \chi_+ =\left(u_{+} ,-i v_{+}\right)^{T}$ for $\beta = \pm 1$
and using the prescriptions described below Eq.~(\ref{Dirac-eq-2*2})
\{see also above
Eqs.~(S7) and (S8) within the SM \cite{SM}\}
write down the corresponding spinors
for $\mathbf{K}_{-}$ point.

\subsection{Equations for the spectrum on the ribbon}

All four solutions for  $\mathbf{K}_{\pm}$ points and
$\beta = \pm 1$ have to satisfy the zigzag boundary conditions
(\ref{boundary-zigzag}) leading to  linear systems of two equations for
the constants $C_{+1}$, $C_{+2}$ and $C_{-1}$, $C_{-2}$.
Equating the system determinants
to zero, one obtains the following transcendental secular equations:
\begin{widetext}
(i) for $\mathbf{K}_+$ point and $\beta = +1$
\begin{equation}
\label{beta=+1_K+point}
\begin{aligned}
\frac{ \left(\delta+\varepsilon\right)   \text{Ai}\left(\frac{\delta ^2-\varepsilon ^2+2 \varepsilon kl  }{(4\varepsilon^2) ^{1/3}}\right)+\frac{2\varepsilon}{(4\varepsilon^2) ^{1/3}} \text{Ai}' \left(\frac{\delta ^2-\varepsilon ^2+2 \varepsilon kl  }{(4\varepsilon^2) ^{1/3} }\right) }{ \left(\delta+\varepsilon\right)   \text{Bi}\left(\frac{\delta ^2-\varepsilon ^2+2 \varepsilon kl  }{(4\varepsilon^2) ^{1/3} }\right)+\frac{2\varepsilon}{(4\varepsilon^2) ^{1/3}} \text{Bi}' \left(\frac{\delta ^2-\varepsilon ^2+2 \varepsilon kl  }{ (4\varepsilon^2) ^{1/3} }\right) } =\frac{  \left(\delta-\varepsilon\right)   \text{Ai}\left(\frac{\delta ^2-\varepsilon ^2+2 \varepsilon \left(W/l+kl\right)}{ (4\varepsilon^2) ^{1/3} }\right)+\frac{2\varepsilon}{(4\varepsilon^2) ^{1/3}} \text{Ai}' \left(\frac{\delta ^2-\varepsilon ^2+2 \varepsilon \left(W/l+kl\right)  }{(4\varepsilon^2) ^{1/3} }\right) }{ \left(\delta-\varepsilon\right)   \text{Bi}\left(\frac{\delta ^2-\varepsilon ^2+2 \varepsilon \left(W/l+kl\right)  }{ (4\varepsilon^2) ^{1/3} }\right)+\frac{2\varepsilon}{(4\varepsilon^2) ^{1/3}} \text{Bi}' \left(\frac{\delta ^2-\varepsilon ^2+2 \varepsilon \left(W/l+kl\right)  }{ (4\varepsilon^2) ^{1/3} }\right) },
\end{aligned}
\end{equation}
(ii) $\mathbf{K}_+$ point and  $\beta=-1$
\begin{equation}
\label{beta=-1_K+point}
\begin{aligned}
\frac{ \left(\delta-\varepsilon\right)   \text{Ai}\left(\frac{\delta ^2-\varepsilon ^2-2 \varepsilon kl  }{(4\varepsilon^2) ^{1/3}}\right)-\frac{2\varepsilon}{(4\varepsilon^2) ^{1/3}}\text{Ai}' \left(\frac{\delta ^2-\varepsilon ^2-2 \varepsilon kl  }{(4\varepsilon^2) ^{1/3}}\right) }{ \left(\delta-\varepsilon\right)   \text{Bi}\left(\frac{\delta ^2-\varepsilon ^2-2 \varepsilon kl  }{(4\varepsilon^2) ^{1/3}}\right)-\frac{2\varepsilon}{(4\varepsilon^2) ^{1/3}}  \text{Bi}' \left(\frac{\delta ^2-\varepsilon ^2-2 \varepsilon kl  }{(4\varepsilon^2) ^{1/3}}\right) } =\frac{  \left(\delta+\varepsilon\right)   \text{Ai}\left(\frac{\delta ^2-\varepsilon ^2-2 \varepsilon \left(W/l+kl\right)}{(4\varepsilon^2) ^{1/3}}\right)-\frac{2\varepsilon}{(4\varepsilon^2) ^{1/3}} \text{Ai}' \left(\frac{\delta ^2-\varepsilon ^2-2 \varepsilon \left(W/l+kl\right)  }{(4\varepsilon^2) ^{1/3}}\right) }{ \left(\delta+\varepsilon\right)   \text{Bi}\left(\frac{\delta ^2-\varepsilon ^2-2 \varepsilon \left(W/l+kl\right)  }{(4\varepsilon^2) ^{1/3}} \right)-\frac{2\varepsilon}{(4\varepsilon^2) ^{1/3}} \text{Bi}' \left(\frac{\delta ^2-\varepsilon ^2-2 \varepsilon \left(W/l+kl\right)  }{(4\varepsilon^2) ^{1/3}}\right) },
\end{aligned}
\end{equation}
(iii) for $\mathbf{K}_-$ point and $\beta=+1$
\begin{equation}
\label{beta=+1_K-point}
\begin{aligned}
\frac{   \left(\delta-\varepsilon\right)   \text{Ai}\left(\frac{\delta ^2-\varepsilon ^2+2 \varepsilon kl  }{(4\varepsilon^2) ^{1/3}}\right)+\frac{2\varepsilon}{(4\varepsilon^2) ^{1/3}} \text{Ai}' \left(\frac{\delta ^2-\varepsilon ^2+2 \varepsilon kl  }{(4\varepsilon^2) ^{1/3}}\right) }{ \left(\delta-\varepsilon\right)   \text{Bi}\left(\frac{\delta ^2-\varepsilon ^2+2 \varepsilon kl  }{(4\varepsilon^2) ^{1/3}}\right)+\frac{2\varepsilon}{(4\varepsilon^2) ^{1/3}} \text{Bi}' \left(\frac{\delta ^2-\varepsilon ^2+2 \varepsilon kl  }{(4\varepsilon^2) ^{1/3} }\right) }= \frac{  \left(\delta+\varepsilon\right)   \text{Ai}\left(\frac{\delta ^2-\varepsilon ^2+2 \varepsilon \left(W/l+kl\right)  }{(4\varepsilon^2) ^{1/3}}\right)+\frac{2\varepsilon}{(4\varepsilon^2) ^{1/3}} \text{Ai}' \left(\frac{\delta ^2-\varepsilon ^2+2 \varepsilon \left(W/l+kl\right)  }{(4\varepsilon^2) ^{1/3}}\right) }{ \left(\delta+\varepsilon\right)   \text{Bi}\left(\frac{\delta ^2-\varepsilon ^2+2 \varepsilon \left(W/l+kl\right)  }{(4\varepsilon^2) ^{1/3}}\right)+\frac{2\varepsilon}{(4\varepsilon^2) ^{1/3}} \text{Bi}' \left(\frac{\delta ^2-\varepsilon ^2+2 \varepsilon \left(W/l+kl\right)  }{(4\varepsilon^2) ^{1/3} }\right) },
\end{aligned}
\end{equation}
(iv) for $\mathbf{K}_-$ point and $\beta=-1$
\begin{equation}
\label{beta=-1_K-point}
\begin{aligned}
\frac{   \left(\delta+\varepsilon\right)   \text{Ai}\left(\frac{\delta ^2-\varepsilon ^2-2 \varepsilon kl  }{(4\varepsilon^2) ^{1/3}}\right)-\frac{2\varepsilon}{(4\varepsilon^2) ^{1/3}} \text{Ai}' \left(\frac{\delta ^2-\varepsilon ^2-2 \varepsilon kl  }{(4\varepsilon^2) ^{1/3}}\right) }{ \left(\delta+\varepsilon\right)   \text{Bi}\left(\frac{\delta ^2-\varepsilon ^2-2 \varepsilon kl  }{(4\varepsilon^2) ^{1/3} }\right)-\frac{2\varepsilon}{(4\varepsilon^2) ^{1/3}} \text{Bi}' \left(\frac{\delta ^2-\varepsilon ^2-2 \varepsilon kl  }{(4\varepsilon^2) ^{1/3}}\right) }= \frac{  \left(\delta-\varepsilon\right)   \text{Ai}\left(\frac{\delta ^2-\varepsilon ^2-2 \varepsilon \left(W/l+kl\right)  }{(4\varepsilon^2) ^{1/3}}\right)-\frac{2\varepsilon}{(4\varepsilon^2) ^{1/3}} \text{Ai}' \left(\frac{\delta ^2-\varepsilon ^2-2 \varepsilon \left(W/l+kl\right)  }{(4\varepsilon^2) ^{1/3}}\right) }{ \left(\delta-\varepsilon\right)   \text{Bi}\left(\frac{\delta ^2-\varepsilon ^2-2 \varepsilon \left(W/l+kl\right)  }{(4\varepsilon^2) ^{1/3}}\right)-\frac{2\varepsilon}{(4\varepsilon^2) ^{1/3}} \text{Bi}' \left(\frac{\delta ^2-\varepsilon ^2-2 \varepsilon \left(W/l+kl\right)  }{(4\varepsilon^2) ^{1/3}}\right) }.
\end{aligned}
\end{equation}
\end{widetext}

\subsection{Spectrum on the half-plane}
\label{sec:half-critical}

The semi-infinite geometry case corresponds to the
limit $W \to \infty$. Consequently,  Eqs.~(\ref{beta=+1_K+point}), (\ref{beta=-1_K+point}),
(\ref{beta=+1_K-point}) and (\ref{beta=-1_K-point}) are greatly simplified
and for $\delta=0$ reduce to
Eqs.~(\ref{beta=+1_K+point-semi}), (\ref{beta=-1_K+point-semi}),
(\ref{beta=+1_K-point-semi}) and (\ref{beta=-1_K-point-semi}).
Specifically, the last equations follow
from utilizing the exponentially divergent asymptotic of
$\text{Bi}(x)$ as $x \to +\infty$ [see Eqs.~(9.7.7) and (9.7.8) of \cite{NIST-book}],
when the right hand sides of Eqs.~(\ref{beta=+1_K+point}) for $\varepsilon > 0$, (\ref{beta=-1_K+point}) for $\varepsilon < 0$,
(\ref{beta=+1_K-point}) for $\varepsilon > 0$, and (\ref{beta=-1_K-point})
for $\varepsilon < 0$ tend to 0.

The hole-like solutions, $\varepsilon < 0$, for $\beta = +1$
and electron-like, $\varepsilon > 0$, for $\beta = -1$ in
Eqs.~(\ref{beta=+1_K+point}), (\ref{beta=-1_K+point}),
(\ref{beta=+1_K-point}) and (\ref{beta=-1_K-point}) cannot be simply
considered in the limit $W \to \infty$. They
correspond to the case of the large negative argument of
$\text{Ai}(-x)$ [see Eq.~(\ref{Airy-expand})] and $\text{Bi}(-x)$
[see Eq.~(9.7.11) of \cite{NIST-book}], where the Airy functions
have oscillatory behavior. Thus to analyze the $W \to \infty$ limit, one may return
to the solutions (\ref{solutions-Airy}). Considering,
for example, the case $\beta =+1$, one can see that for $\varepsilon \leq 0$
the boundary condition (\ref{boundaryZ-W}) for $\varepsilon < 0$
can only be satisfied by the trivial solution $C_{+1} = C_{+2} =0$.
The wave functions with all energies $\varepsilon < 0$ vanish and corresponding
states disappear from the Hilbert space.
The case $\varepsilon = 0$ for $\beta = \pm 1$ is treated by returning to
the original system (\ref{chi-tilde}). One can check that the normalizable
solutions are absent in the semi-infinite geometry. This agrees with
Ref.~\cite{Cheng2013Few}.

The resulting equations for the spectra on the half-plane
for $\Delta =0$
are the following:\\
(i) for $\mathbf{K}_+$ point, $\beta =+1$, and $\varepsilon>0$
\begin{equation}
\label{beta=+1_K+point-semi}
 \hspace*{-0.2cm}
\text{Ai}\left(\frac{-\varepsilon ^2+2 \varepsilon kl  }{(4\varepsilon^2) ^{1/3}}\right)+\left( \frac{2}{\varepsilon^2} \right)^{1/3} \text{Ai}' \left(\frac{-\varepsilon ^2+2 \varepsilon kl  }{(4\varepsilon^2) ^{1/3} }\right) =0,
\end{equation}
(ii) for $\mathbf{K}_+$ point, $\beta =-1$, and $\varepsilon<0$
\begin{equation}
\label{beta=-1_K+point-semi}
 \hspace*{-0.2cm}
\text{Ai}\left(\frac{-\varepsilon ^2-2 \varepsilon kl  }{(4\varepsilon^2) ^{1/3}}\right)+\left( \frac{2}{\varepsilon^2} \right)^{1/3} \text{Ai}' \left(\frac{-\varepsilon ^2-2 \varepsilon kl  }{(4\varepsilon^2) ^{1/3}}\right) =0,
\end{equation}
(iii) for $\mathbf{K}_-$ point, $\beta =+1$, and $\varepsilon>0$
\begin{equation}
\label{beta=+1_K-point-semi}
 \hspace*{-0.2cm}
\text{Ai}\left(\frac{-\varepsilon ^2+2 \varepsilon kl  }{(4\varepsilon^2) ^{1/3}}\right)-\left( \frac{2}{\varepsilon^2} \right)^{1/3}  \text{Ai}' \left(\frac{-\varepsilon ^2+2 \varepsilon kl  }{(4\varepsilon^2) ^{1/3}}\right) =0,
\end{equation}
(iv) for $\mathbf{K}_-$ point, $\beta =+1$, and $\varepsilon<0$
\begin{equation}
\label{beta=-1_K-point-semi}
 \hspace*{-0.2cm}
\text{Ai}\left(\frac{-\varepsilon ^2-2 \varepsilon kl  }{(4\varepsilon^2) ^{1/3}}\right)-\left( \frac{2}{\varepsilon^2} \right)^{1/3}  \text{Ai}' \left(\frac{-\varepsilon ^2-2 \varepsilon kl  }{(4\varepsilon^2) ^{1/3}}\right)=0.
\end{equation}

\begin{figure}[!ht]
\includegraphics[width=.42\textwidth]{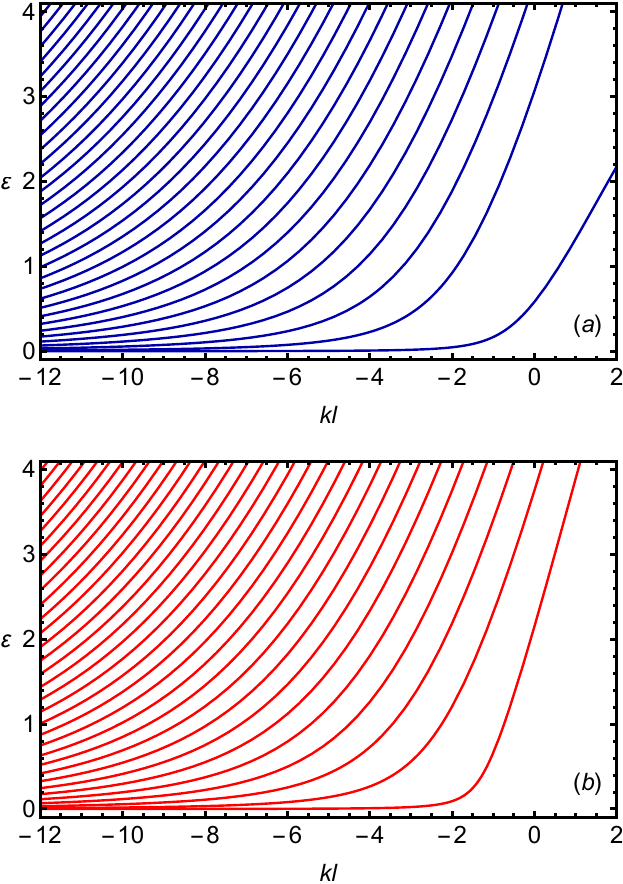}
\caption{The energy spectra $\varepsilon(k)$ in the critical regime,
$\beta =1$, near the zigzag edge of graphene's half-plane for the gapless,
$\delta =0$ case. (a) For $\mathbf{K}_+$ valley, (b) For $\mathbf{K}_-$
valley.  }
\label{fig:critical-regime}
\end{figure}

In Fig.~\ref{fig:critical-regime}
we present numerical solutions of Eqs.~(\ref{beta=+1_K+point-semi}) and (\ref{beta=+1_K-point-semi}) for $\varepsilon (k l) >0$, which describe electron-like levels in the critical regime.
The hole-like solutions that were present for $\beta < 1$
and collapsing towards the $\varepsilon = 0$ level in the $\beta \to 1$ limit
are absent because there are no normalizable  solutions of the original
system in the $\beta = \pm 1$ cases. It is shown
in Ref.~\cite{Cheng2013Few} that the critical solutions, $\beta = \pm 1$,
are not bound states.

One can easily see that  Eqs.~(\ref{beta=+1_K+point-semi}) and (\ref{beta=+1_K-point-semi}) for $kl =0$ reduce to Eqs.~(\ref{Airy-K+})
and (\ref{Airy-K-}), respectively. Accordingly, the  energies of the Landau
levels at the edge in Fig.~\ref{fig:collapse-edge} (for $\beta =+1$) and Fig.~\ref{fig:critical-regime} (for $kl =0$) are in agreement
and tend to the values given by Eqs.~(\ref{e-levels-set-beta=1})
and (\ref{e-levels-set-beta=1-approx}).  As discussed in the previous section,
these electron-like levels do not collapse at the edge, but now we may also follow
their behavior in the bulk. One can see in Fig.~\ref{fig:critical-regime} that
these levels become denser and approach each other asymptotically
as $k l \to - \infty$. This conclusion is confirmed by
studying analytically  Eqs.~(\ref{beta=+1_K+point-semi}) and (\ref{beta=+1_K-point-semi}) in the limit $-kl = L/l$ for $L \gg l$
with $L$ being the distance from the edge.
The derivation follows the approach to Eq.~(\ref{e-levels-set-beta=1-approx}) and
relies on the asymptotic expressions given by Eq.~(\ref{Airy-expand}).
It gives that the distance between the levels in the bulk
is $\sim l^3/L^3$ for $L \gg l.$
This implies that {\it there is no Landau level collapse}
of the electron-like states in the semi-infinite geometry for $\beta =1$.
This conclusion is, however, correct in the formal mathematical sense.
The presence of disorder inevitably causes the broadening of Landau levels and,
accordingly, the levels of a finite width would anyway merge.

\subsection{Specific of the Landau level collapse on the zigzag ribbon}

The investigation of the energy levels on the
ribbon was done either on the lattice fully numerically
\cite{Lukose2007PRL,Roslyak2010PTRS,Ostahie2015PRB}
or by solving transcendental equation as in
Secs.~S2 and S3 within the SM \cite{SM}.
Although it allows to observe that
the levels in certain regions of the ribbon become denser than in
others, one cannot conclusively demonstrate that they collapse in a manner akin to what occurs in the case of an infinite system [see Eq.~(\ref{LL-collapse})].
In this respect, the analysis of the level behavior in the half-plane
geometry in Secs.~\ref{sec:zigzag-semi} and \ref{sec:armchair-semi}
is more convincing because it allows one to distinguish the level
behavior in the $|\beta| \to 1$ limit.

Indeed, we have discovered that in the case of edges with bulk states situated on the left side, as the limit $\beta \to 1$ is approached, the hole-like levels collapse across the entire semi-plane.
However, the point of the collapse, $\varepsilon =0$ and $\beta =1$,
does not belong to the bound state spectrum.
Conversely, the electron-like levels do not collapse at all.
Their energies near the edge tend to the different values and
deep within the bulk, when $-kl = L/l \gg 1$, the levels approach each
other asymptotically with the distance between them $\sim l^3/L^3$.

\begin{figure}[!ht]
\includegraphics[width=.42\textwidth]{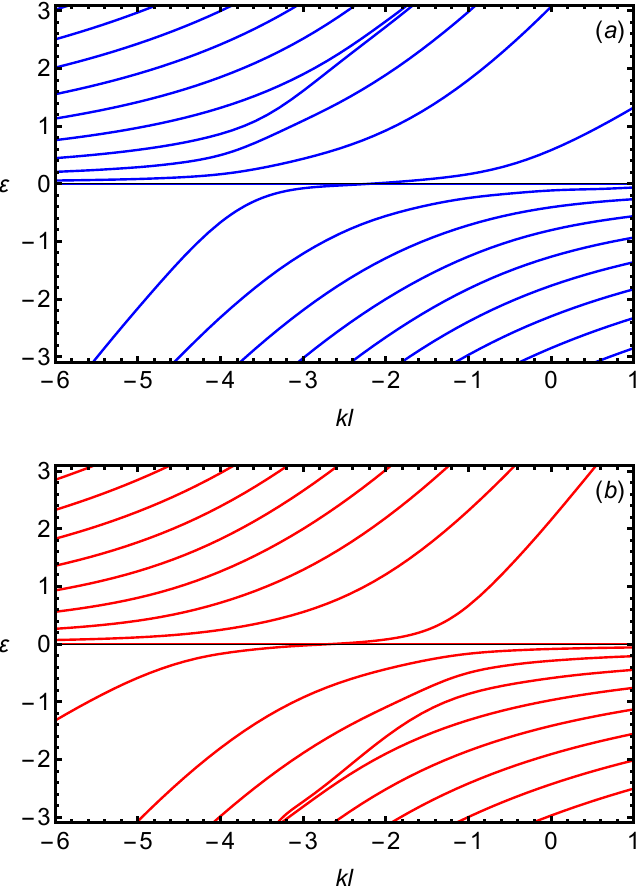}
\caption{The energy spectra $\varepsilon(k)$ in the critical regime,
$\beta =1$, of the zigzag ribbon of width $W = 5l$ for the gapless,
$\delta =0$ case obtained by the numerical
Eqs.~(\ref{beta=+1_K+point}) and (\ref{beta=+1_K-point}).
(a) For $\mathbf{K}_+$ valley, (b) For $\mathbf{K}_-$ valley.  }
\label{fig:critical-regime-ribbon}
\end{figure}

Now we discuss the level behavior in the ribbon geometry.
The critical regime, $\beta =1$,
is shown in Fig.~\ref{fig:critical-regime-ribbon}
for a ribbon of width $W = 5 l$, which is chosen to be smaller than for the
rest of the figures for a better level resolution.
First of all we observe that {\it there is no Landau level collapse}
on the zigzag ribbons. While the hole-like levels were collapsing
in the semi-infinite geometry, now they only become denser near the
$k l =0$ edge.

This result can be qualitatively explained as follows: as observed earlier, in the semi-infinite geometry, the collapse of Landau levels happens when the center of the hole orbit remains in the bulk [$y>0$, see Eq.~(\ref{orbital-generic})].
For $\beta =1$ this is possible for hole-like states,
but not for the electron-like ones. The presence of the other edge
does not allow the hole orbit to remain inside the ribbon.
We also note the interchange of electron- and hole-like levels
at the opposite edge, as expected based on symmetry arguments.

To extend the analysis of Sec.~\ref{sec:half-critical}
of the level behavior in the bulk to the case of the ribbon,
we consider
Eqs.~(\ref{beta=+1_K+point}), (\ref{beta=-1_K+point}),
(\ref{beta=+1_K-point}) and (\ref{beta=-1_K-point}) from Sec.~\ref{sec:critical}.
We obtain that for $\beta =1$ the distance between electron-like levels
near $-kl = W/l$ edge is $O (l^3/W^3)$ for $W >> l$.
This confirms our statement that {\it there is no Landau level collapse} on the ribbons in a sense that all levels do not collapse to the lowest one. This conclusion is, however, correct only mathematically,
because if the ribbon is wide enough the Landau levels of a finite
width would anyway merge.
On the other hand, we obtained that the correction to the
non-collapsing solutions (\ref{e-levels-set-beta=1}) is
$O \left( e^{- 4 \sqrt{2 |\varepsilon|}/3 (W^3/l^3)} \right)$.


\section{Half-plane with the armchair edge}
\label{sec:armchair-semi}


As was discussed in Sec.~\ref{sec:armchair},
the armchair boundary conditions (\ref{boundary-armchair})
admix the solutions for $\mathbf{K}_{\pm}$ valleys. Thus
in addition to the solution  (\ref{sol-tilde-chi1})
and (\ref{sol-tilde-chi2}) for $\mathbf{K}_+$ valley
one needs the corresponding solutions
(S7)  and (S8) from the SM \cite{SM} for $\mathbf{K}_-$ valley.
The equation (S13)  for the spectrum of the armchair ribbon
undergoes significant simplification in the case of semi-infinite geometry,
where $W \to \infty$. As discussed in Sec.~\ref{sec:zigzag-semi} for
the half-plane with the zigzag edge, one may set $C_{+2} = C_{-2} =0$.

Now the boundary conditions (\ref{boundaryA-W})
at $x = W \to \infty$ are automatically satisfied as a result of the
asymptotic of $U(a,x \to \infty)$
and the remaining boundary conditions
(\ref{boundaryA-0}) at $x=0$ result in the following
system of equations:
\begin{equation}
\label{system-armchair_half}
\begin{aligned}
& C_{+1} \left[ \gamma U \left(a-1,\sqrt{2} \zeta(0) \right)-
\kappa_{+} U \left(a,\sqrt{2} \zeta(0) \right)\right]\\
+ & C_{-1} \left[  U \left((a-1,\sqrt{2} \zeta(0) \right)+
\gamma \kappa_{-} U \left(a,\sqrt{2} \zeta(0) \right)\right]=0,\\
& C_{+1} \left[  U \left(a-1,\sqrt{2} \zeta(0) \right)-
\gamma \kappa_{+} U \left(a,\sqrt{2}\zeta(0) \right)\right]\\
+ &C_{-1} \left[ \gamma U(a-1,\sqrt{2}\zeta(0))+ \kappa_{-}
U(a,\sqrt{2}\zeta(0))\right]=0.
\end{aligned}
\end{equation}

The requirement for the system (\ref{system-armchair_half}) to possess nontrivial solution
results in the following equation for the spectrum:
\begin{equation}
\label{spectrum-armchair_half}
 \hspace*{-0.15cm}
U^2 \left(a-1,\sqrt{2} \zeta(0) \right) +
\left(a-\frac{1}{2}\right) U^2 \left(a,\sqrt{2} \zeta(0) \right)=0.
\end{equation}
It determines dimensionless energies
$\varepsilon_\alpha=\varepsilon_n(kl)$
as functions of quantum numbers $\alpha\equiv n, k$
of a half-plane with the armchair edge.

\begin{figure}[!ht]
\includegraphics[width=.42\textwidth]{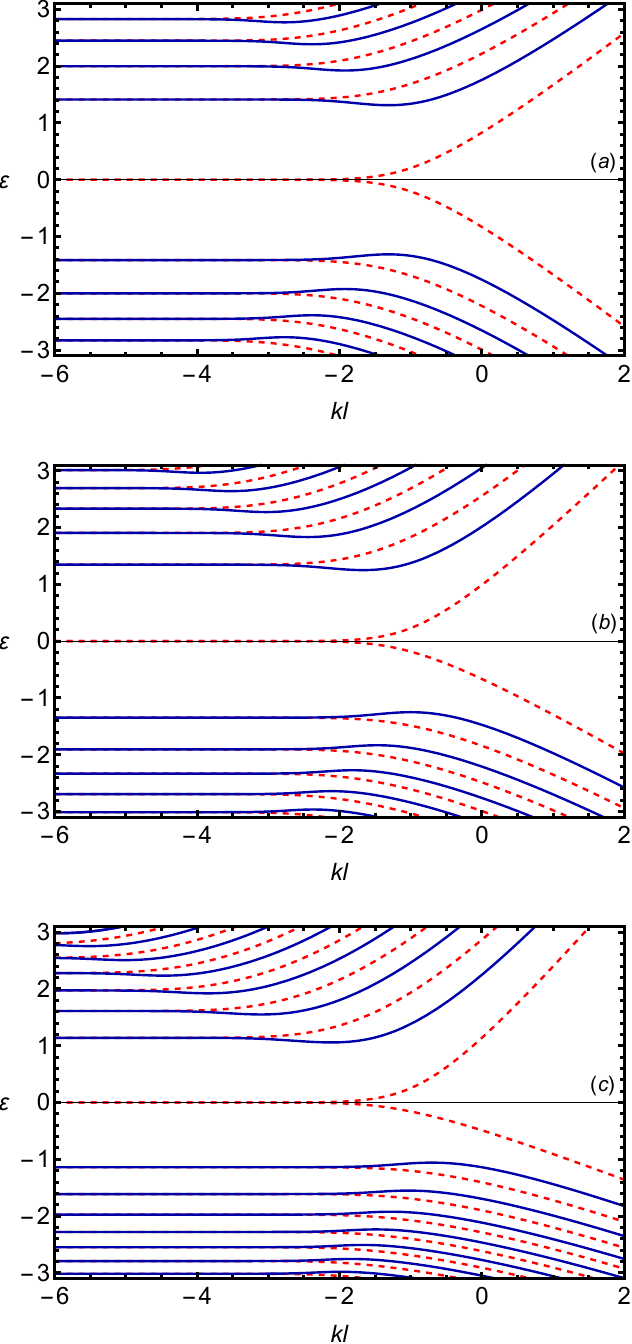}
\caption{The energy spectra $\varepsilon (k)$
of the first few Landau levels near the armchair edge of graphene
for the gapless ($\delta=0$) case. (a) $\beta =0$; (b)
$\beta =0.25$;  and (c) $\beta =0.5$.
The solid (blue) and the dashed (red) lines are solely used
for a better resolution of the degenerate in the bulk solutions.
}
\label{fig:armchair-spectrum}
\end{figure}

\begin{figure}[!ht]
\includegraphics[width=.42\textwidth]{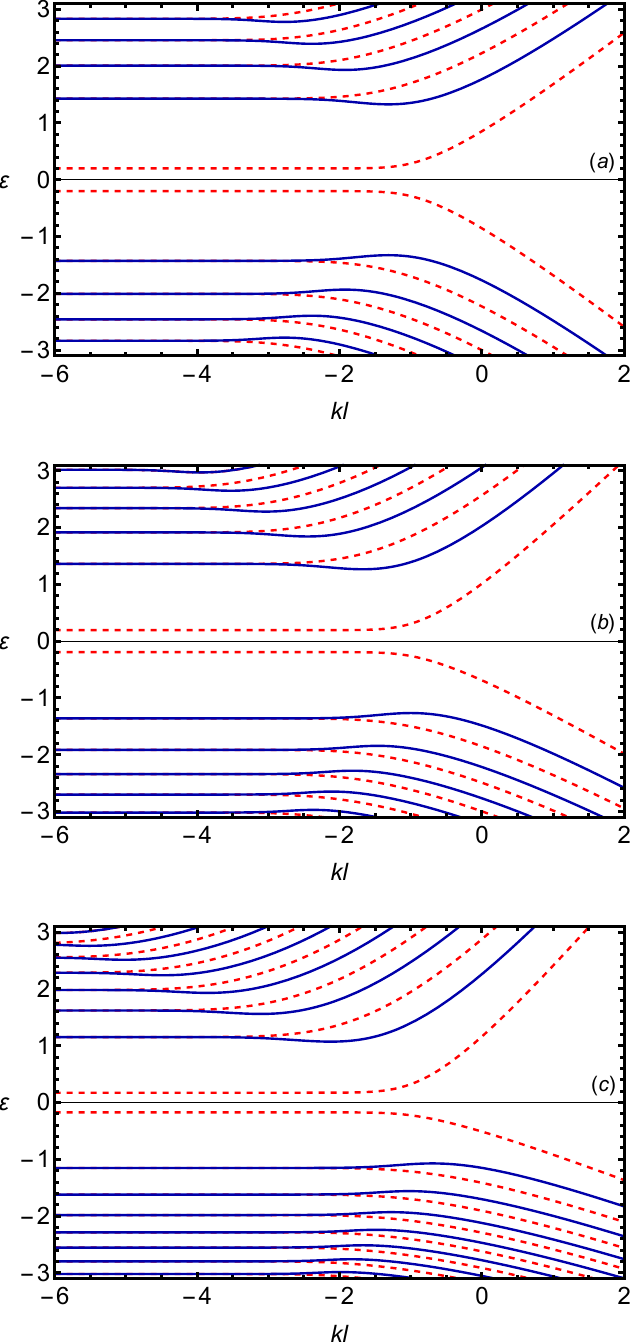}
\caption{The energy spectra $\varepsilon (k)$
of the first few Landau levels near the armchair edge of graphene for the gapped ($\delta=0.2$) case.
The panels (a), (b), and (c) are for the same values of $\beta$ as in  Fig.~\ref{fig:armchair-spectrum}.
}
\label{fig:armchair-gapped-spectrum}
\end{figure}

The corresponding spectra are computed numerically and
presented for the gapless and gapped cases in Figs.~\ref{fig:armchair-spectrum}
and \ref{fig:armchair-gapped-spectrum}, respectively, where
the dimensionless energy $\varepsilon(k)$ is plotted.
Let us recapitulate the main specific features of these solutions.
Since a half-plane geometry is considered, for a finite $\beta$
the energies $\varepsilon(k)$ tend to constant values in the bulk
as $k l \to -\infty$.
The levels, whose degeneracy was lifted by the edge become degenerate in the bulk.
The armchair edge does not support the dispersionless mode.
However, one of the solutions when approaching the
edge shows a nonmonotonous behavior of the energy.
This leads to a change in the sign of the drift velocity
\cite{Wang2011EPJB,Deplace2010PRB}.

\subsection{Zero electric field limit}

In the absence of electric field, $\beta = 0$,
Eq.~(\ref{spectrum-armchair_half}) preserves its
form, but with $a = [1 + \delta^2 - \varepsilon^2]/2$
and $\zeta(0) = kl$ which agrees with
the equation studied in Refs.~\cite{Gusynin2008PRB,Gusynin2008FNT}.
The corresponding numerical solutions shown in Figs.~\ref{fig:zigzag-spectrum}~(a)
and \ref{fig:zigzag-gapped-spectrum}~(a)
are in agreement with the results presented in Refs.~\cite{Ababnin2006PRL,Brey2006bPRB,Gusynin2008PRB,Gusynin2008FNT}.

\subsection{Finite electric field}
\label{sec:armchar-finiteE-edge}

\begin{figure}[!ht]
\includegraphics[width=.42\textwidth]{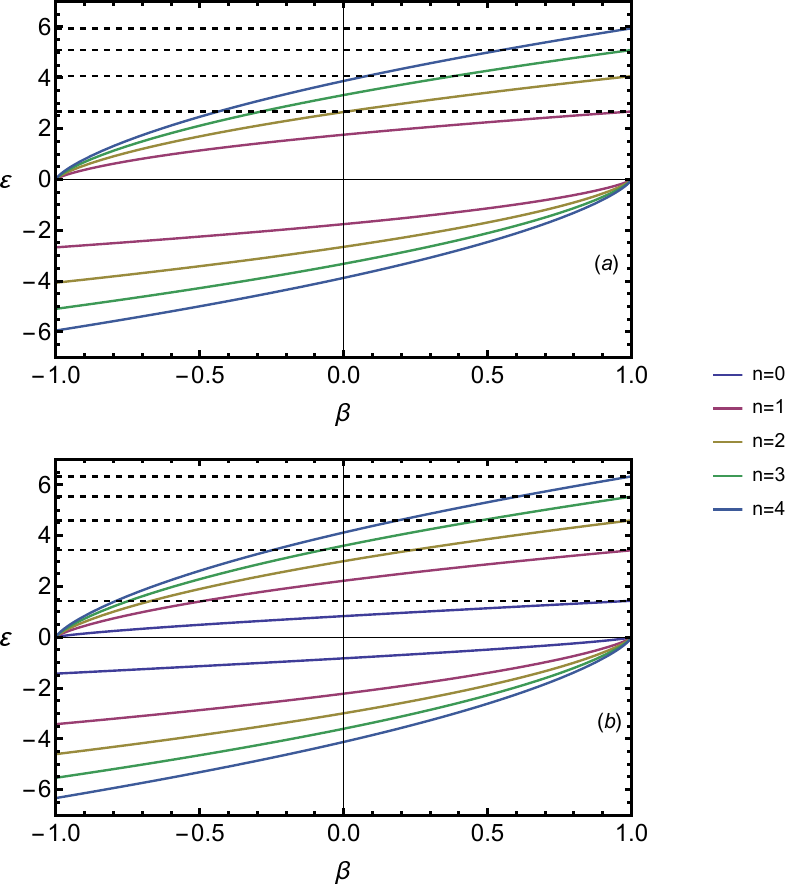}
\caption{The energy spectra $\varepsilon(kl =0)$ at the armchair edge
versus electric field in terms of $\beta = c E /(v_F H)$
for $\delta =0$ for first few Landau levels.
(a) The dashed lines correspond to the energies
given by the upper line of Eq.~(\ref{e-levels-set-beta=1-armchair}).
(b) The dashed lines correspond to the energies
given by the lower line of Eq.~(\ref{e-levels-set-beta=1-armchair}).
}
\label{fig:collapse-edge-armchair}
\end{figure}

As in Sec.~\ref{sec:zigzag-finiteE-edge}, we show
in Fig.~\ref{fig:collapse-edge-armchair} numerical solutions
of Eq.~(\ref{spectrum-armchair_half})
for the energy spectra at the armchair edge, $\varepsilon(kl =0)$,
and $\delta =0$.
These solutions are depicted as functions of $-1 \leq \beta \leq 1$ for the
first few Landau levels. We emphasize that for the armchair edge the solutions for
$\mathbf{K}_\pm$ valleys are not separable and the panels (a) and (b)
are solely used for a better readability.
One can see that the behavior of the level energies
at the edge is rather similar to the case of the zigzag edge.
The electron-like levels  collapse
only for $\beta \to -1$, while for $\beta \to 1$ the levels do not
collapse and their energies tend to the different values.
On the contrary, the hole-like solutions merge and collapse for
$\beta \to 1$, while for $\beta \to -1$ these levels do not collapse.

Similar to the zigzag edge case, it is possible to consider analytically
Eq.~(\ref{spectrum-armchair_half}) at the edge, $k l = 0$.
The spectrum in the $\beta \to 1$ limit reads
\begin{equation}
\label{e-levels-set-beta=1-armchair}
\varepsilon_{n} =
\begin{cases}
\sqrt{2 \varkappa^3_{1,n}},\\
\sqrt{2 \varkappa^3_{2,n}},
\end{cases}
\qquad n =0,1, \ldots,
\end{equation}
where  $\varkappa_{1,n}$ and $\varkappa_{2,n}$  are the roots of the equations
$\text{Ai}\left( -x^2  \right)=0$ and $\text{Ai}' \left( -x^2  \right)=0$,
respectively.
These values of the energies corresponding to the upper and lower lines
of Eq.~(\ref{e-levels-set-beta=1-armchair}) are
shown in Figs.~\ref{fig:collapse-edge-armchair}~(a) and \ref{fig:collapse-edge-armchair}~(b), 
respectively, confirming a
good agreement between the original equation for the spectrum and
its $\beta \to 1$ limit.
Finally, we obtain the following approximation for the
solution (\ref{e-levels-set-beta=1-armchair})
\begin{equation}
\label{e-levels-set-beta=1-armchair-trig}
\varepsilon_{n}^{\mathrm{app}} =
\begin{cases}
\sqrt{3 \pi \left(n-\frac{1}{4} \right)} , &  n =1,2 \ldots, \\
\sqrt{3 \pi \left(n+\frac{1}{4} \right)} , & n =0,1, \ldots \\
\end{cases}
\end{equation}
which shows a good agreement
with the results obtained by numerically solving
the full equation involving the Airy function.

Although the complexity of the corresponding equations makes it impossible to analyze the level behavior for the armchair ribbon as was done for the zigzag ribbon, the present results indicate that there is no Landau level collapse in armchair ribbons.

\section{Summary}
\label{sec:conclusion}

To conclude our main results can be summarized as follows:
\begin{enumerate}

\item[(i)] The Landau level collapse in the restricted geometry
occurs not in the same fashion as in the infinite geometry where
it appears as a sharp transition.

\item[(ii)] In the semi-infinite geometry the hole
(electron)-like Landau levels
collapse as the ratio of electric and magnetic fields, $\beta = c E /v_F H$, reaches the critical value $ +(-) 1$. On the other hand, the energies of the electron
(hole)-like levels near the edge remain different and are given by Eqs.~(\ref{e-levels-set-beta=1-approx})
and (\ref{e-levels-set-beta=1-armchair-trig}) for the zigzag and
armchair edges, respectively. The same levels deeply within the bulk,
for $-kl = L/l \gg 1 $, approach each other asymptotically.

\item[(iii)]
There is no Landau level collapse on the ribbons, because
in contrast to the semi-infinite geometry the orbit center cannot go
to infinity. Instead, the electron (hole)-like levels become denser.
The absence of collapse is, however, valid in a mathematical sense,
because if the ribbon is wide enough the Landau levels of a finite
width would anyway merge forming a band.

\item[(iv)] We derived the transcendental equations describing the Landau level
behavior in the crossed magnetic and in-plane electric fields
on the zigzag and armchair ribbons (see the SM \cite{SM})
with the edges at $kl =0$ and $kl = - W/l$
and in the semi-infinite geometry, $W \to \infty$ [Eqs.~(\ref{spectrum-zigzag+half}), (\ref{spectrum-zigzag-half}), and (\ref{spectrum-armchair_half})].
These equations are analyzed analytically and numerically.

\item[(v)] In particular, the simplified equations
(\ref{beta=+1_K+point}), (\ref{beta=-1_K+point}),
(\ref{beta=+1_K-point}) and (\ref{beta=-1_K-point}) in terms of the
Airy functions describing the critical regime are obtained.

\end{enumerate}

The obtained behavior of the Landau level collapse on the ribbons represents
a particular example of describing systems of a finite size.
As   mentioned in the Introduction, the Landau-level collapse was
already observed experimentally \cite{Singh2009PRB,Gu2011PRL}.
It would be useful to test the specific predictions made in the
present study.

\begin{acknowledgments}
We would like to thank the Armed Forces of Ukraine for providing security to perform this work.
A.A.H acknowledges support from the National Research Foundation of Ukraine Grant  (2020.02/0051) 
``Topological phases of matter and excitations in Dirac materials, Josephson junctions and magnets''.
S.G.Sh and V.P.G. acknowledge the partial financial support from the
National Academy of Sciences of Ukraine, Project No.~0123U102283 in 2023,
and Projects No.~0121U109612 and No.~0122U000887 in 2024.
We would like to express our gratitude to
I.A.~Shovkovy for insightful discussions on numerical methods.

\end{acknowledgments}

\appendix

\section{Wave functions for semi-infinite system}
\label{sec:Appendix-probability}

In deriving Eqs.~(\ref{spectrum-zigzag+half}) and (\ref{spectrum-zigzag-half})
for the spectra we already obtained from the general expressions for
the wave functions $ \Psi_+^T = \left( \psi_{AK_+}, \psi_{BK_+} \right) $
given by Eqs.~(\ref{sol-tilde-chi1}) and (\ref{sol-tilde-chi2})
and $\Psi_-^T = \left(\psi_{BK_-}, \psi_{AK_-} \right)$
given by Eqs.~(S7) and (S8) (see the SM \cite{SM}), respectively.
In the half-plane geometry they reduce to the following ones
\begin{equation}
\label{wave-function-K+}
\begin{split}
& \Psi_+(\mathbf{r},k)=\frac{e^{-ikx}}{\sqrt{2\pi l}} \times \\
&\left( \begin{array}{c}
i C_{+1} \left[ \gamma U \left(a-1,\sqrt{2} \zeta \right)-\kappa_{+}  U \left(a,\sqrt{2} \zeta \right)\right] \\
 C_{+1} \left[ U \left(a-1,\sqrt{2} \zeta \right)-  \gamma \kappa_{+}  U \left(a,\sqrt{2} \zeta \right)\right]
\end{array}\right).
\end{split}
\end{equation}
and
\begin{equation}
\label{wave-function-K-}
\begin{split}
& \Psi_-(\mathbf{r},k)=\frac{e^{-ikx}}{\sqrt{2\pi l}} \times \\
&\left( \begin{array}{c}
 C_{-1} \left[ \gamma U \left(a-1,\sqrt{2} \zeta \right)+\kappa_{-}  U \left(a,\sqrt{2} \zeta \right)\right] \\
i C_{-1} \left[  U \left(a-1,\sqrt{2} \zeta \right)+\gamma \kappa_{-}    U \left(a,\sqrt{2} \zeta \right)\right]
\end{array}\right).
\end{split}
\end{equation}
Then the  normalization condition for a given valley $\mathbf{K}_{\eta}$ is defined as follows:
\begin{equation}
\begin{split}
& \int_{-\infty}^{+\infty}dx\int_{0}^{+\infty}dy \Psi_{\eta,n}^{(\alpha)\dagger}(\mathbf{r},k) \Psi_{\eta',n'}^{(\alpha)}(\mathbf{r},k')\\
& =\delta \left(k-k'\right) \delta_{\eta \eta'}  \delta_{n n'}
\end{split}
\end{equation}
where $n$ is the Landau level index, $\alpha  =e,h$ is the electon-like ($\varepsilon >0$)
or hole-like ($\varepsilon <0$) levels.
Hence, the normalization constants $C_{\pm 1}$ for the wave functions (\ref{wave-function-K+})
and (\ref{wave-function-K-}) are
\begin{equation}
\begin{split}
 &\left| C_{\pm 1} \right|^{-2}  = \frac{2}{1+\sqrt{1-\beta^2} } \int_{0}^{+\infty}d\xi \\
&\left\{ \left[ U \left(a-1,\sqrt{2} \zeta \right) \right]^2
 \mp 2\beta \kappa_{\pm}U \left(a-1,\sqrt{2} \zeta \right) \right. \\
&  \left. \times U \left(a,\sqrt{2} \zeta \right)+
\left[ \kappa_{\pm} U \left(a,\sqrt{2} \zeta \right) \right]^2 \right\}.
\end{split}
\end{equation}
Accordingly, this enables us to consider the squared modulus of the wave function
$\Psi_{\eta,n}^{(\alpha)}(\mathbf{r},k)$ as defined by Eq.~(\ref{probability}).

\section{Analytical study of the spectrum for semi-infinite system with zigzag
edge, $kl=0$}
\label{sec:Appendix-zigzag}

For small values of $\beta$, Eqs.~(\ref{spectrum-zigzag+half}) and (\ref{spectrum-zigzag-half}) can be solved by expanding the parabolic cylinder
function $U(a,z)$ in the argument $z$ [see Eq. (19.3.5) in Ref.\cite{Abramowitz.book}]. In this case, we have
\begin{equation}
\sqrt{2} \zeta(0) \Big|_{kl=0}=\sqrt{2} \frac{\beta \varepsilon}{\left(1-\beta^2 \right)^{3/4}}=\sqrt{2} \beta \varepsilon^{(0)} +o(\beta),\\
\end{equation}
and
\begin{equation}
\begin{split}
\kappa_{-} \kappa_{+} & =\frac{\delta^2\left( 1-\beta^2\right) -\varepsilon^2}{2\left(1-\beta^2 \right)^{3/2}} \\
& =\frac{1}{2}\left( \delta^2 -(\varepsilon^{(0)})^2\right)-\varepsilon^{(0)} \varepsilon^{(1)} \beta +o(\beta),
\end{split}
\end{equation}
where $\varepsilon= \sum_{s=0}^{\infty} \varepsilon^{(s)} \beta^s$,
$\varepsilon = \varepsilon(kl =0)$.

Then expanding the solutions for the $\mathbf{K}_{+}$ valley
in the vicinity the $\beta=0$ solutions, $\varepsilon_{+,0}^{(0)}=-\delta$ and $(\varepsilon_{+,n}^{(0)})^2-\delta^2=4n$ with $n=1,2, \ldots$
\{see Eq.~(S17) in the SM \cite{SM}\}
we obtain the following set of equations for $\varepsilon^{(1)}$:
\begin{equation}
 \frac{\beta }{\Gamma\left(\frac{1}{2}\right)}  =  \varepsilon^{(1)}_{+,0} \beta,
\end{equation}
\begin{equation}
\begin{split}
 \frac{\beta +2 \beta \varepsilon^{(0)}_{+,n}  \left( \delta+ \varepsilon^{(0)}_{+,n} \right) }{\Gamma\left(\frac{1}{2}-n\right)}& = \\
 \left( \delta + \varepsilon^{(0)}_{+,n} \right)
(-1)^{n-1} & \Gamma(n)
\left(- \frac{1}{2}\varepsilon^{(0)}_{+,n} \varepsilon^{(1)}_{+,n} \beta \right)
\end{split}
\end{equation}
Hence, we arrive at Eq.~(\ref{sol-K+beta-small}).

Similarly,  for the $\mathbf{K}_{-}$ valley expanding in the vicinity of the solutions for $\beta=0$: $\varepsilon_{-,n}^2-\delta^2=2(2n+1)$
with $n=0,1, \ldots$ \{see Eq.~(S18) within the SM \cite{SM}\}
we obtain the following set of equations for $\varepsilon^{(1)}$
\begin{equation}
\begin{split}
& \frac{2\beta \varepsilon^{(0)}_{-,n} \left(\delta^2-(\varepsilon^{(0)}_{-,n})^2\right) - \beta \left( \delta- \varepsilon^{(0)}_{-,n} \right) }{4 \Gamma\left(\frac{1}{2}-n\right)} \\
& = (-1)^{n} \Gamma(n+1) \left(- \frac{1}{2}\varepsilon^{(0)}_{-,n} \varepsilon^{(1)}_{-,n} \beta \right)
\end{split}
\end{equation}
and finally arrive at Eq.~(\ref{sol-K-beta-small}).

Let us consider electronic levels with $\varepsilon > 0$ and $\delta =0$
in the limit $\beta \rightarrow 1$. We rewrite the
arguments of the parabolic cylinder functions that enter
Eqs.~(\ref{spectrum-zigzag+half}) and (\ref{spectrum-zigzag-half})
as follows:
\begin{widetext}
\begin{equation}
\label{U-rewritten}
\begin{rcases}
 U \left(a , \sqrt{2} \zeta(0) \Big|_{kl=0}\right) & \\
 U \left(a-1 ,\sqrt{2} \zeta(0) \Big|_{kl=0}\right)&
\end{rcases} =
U\left( -\frac{\lambda}{2}-\frac{\nu^2}{2} , \sqrt{2} \beta \nu \right)
= U\left( -\frac{\mu_\lambda^2}{2} , \sqrt{2} \beta_\lambda \mu_\lambda \right),
\end{equation}
where the following notations were introduced
$\nu= \varepsilon \left(1-\beta^2\right)^{-3/4}$,
$\mu_{\lambda}= \nu\sqrt{1+\frac{\lambda}{\nu^2}}$ and $\beta_{\lambda}=\beta \left(1+\frac{\lambda}{\nu^2}\right)^{-\frac{1}{2}}$ with $\lambda =\mp 1$.

According to Ref.~\cite{NIST-book}, Eq.~(12.10.35) [see also Eq.~(3.1)
in \cite{Temme2000JCAM}] for $a =-\mu^2/2 <0$, $-2 \sqrt{-a} < \sqrt{2}\mu \beta < \infty$ and for large positive real values
of $\mu$ one can use the
asymptotic expansion of the parabolic cylinder function
in terms of Airy functions $\text{Ai}(x)$
that for $\beta \in [-1+\epsilon,\infty )$ converges uniformly,
\begin{equation}
\label{Temme-expansion}
U\left( -\frac{1}{2} \mu^2 , \sqrt{2} \beta \mu \right) = 2\pi^{\frac{1}{2} } \mu^{\frac{1}{3} }g (\mu) \phi (\zeta) \left[ \text{Ai} \left( \mu^{\frac{4}{3} } \zeta \right) A_{\mu} (\zeta)+\frac{\text{Ai}' \left( \mu^{\frac{4}{3} } \zeta \right) }{\mu^{\frac{8}{3} }}B_{\mu} (\zeta) \right],
\end{equation}
where
\begin{equation}
\label{short-notations}
\begin{split}
& \zeta=-\left( \frac{3}{2} \eta \right)^{\frac{2}{3}}, \qquad
\eta =\frac{1}{2} \text{arccos} (\beta) -\frac{1}{2} \beta \sqrt{1-\beta^2}, \qquad
g(\mu)    \sim  2^{-\frac{1}{4} \mu^2 -\frac{1}{4}} e^{-\frac{1}{4} \mu^2 } \mu^{\frac{1}{2} \mu^2-\frac{1}{2} }\left( \sum_{s=0}^{\infty} \frac{g_s}{\mu^{2s}} \right)^{-1}, \\
& \phi (\zeta)=\left( \frac{-\zeta}{1-\beta^2} \right)^{ \frac{1}{4} }, \qquad
A_{\mu}(\zeta) \sim \sum_{s=0}^{\infty} \frac{a_s(\zeta)}{\mu^{4s}}, \qquad
B_{\mu}(\zeta)\sim \sum_{s=0}^{\infty} \frac{b_s(\zeta)}{\mu^{4s}}
\end{split}
\end{equation}
and the coefficients $a_s(\zeta)$, $b_s(\zeta)$ and $g_{s}$ are given in
Eqs.~(3.9) and (2.8), respectively, from Ref.~\cite{Temme2000JCAM}.
Taking the leading term of the expansion we obtain
\begin{equation}
\label{U-ratio}
\frac{U\left( -\frac{1}{2} \mu_{+}^2 , \sqrt{2} \beta_{+} \mu_{+} \right)}{U\left( -\frac{1}{2} \mu_{-}^2 , \sqrt{2} \beta_{-} \mu_{-} \right) } = \frac{g(\mu_{+})}{g(\mu_{-})} \frac{\mu^{\frac{1}{3}}_{+}  \phi (\zeta_{+}) \text{Ai} \left( \mu^{\frac{4}{3}}_{+} \zeta_{+} \right) }{\mu^{\frac{1}{3}}_{-}  \phi (\zeta_{-}) \text{Ai} \left( \mu^{\frac{4}{3}}_{-} \zeta_{-} \right)}
\underset{\beta \rightarrow 1}{\sim} \frac{\nu}{\sqrt{2}}
\frac{\text{Ai} \left( -\left(\frac{\varepsilon^2}{2} \right)^{\frac{2}{3}}  \left(1+  \frac{\sqrt{2} \sqrt{1-\beta} }{ \varepsilon^2 } \right)\right) }{ \text{Ai} \left( -\left(\frac{\varepsilon^2}{2} \right)^{\frac{2}{3}}  \left(1 - \frac{\sqrt{2} \sqrt{1-\beta} }{ \varepsilon^2 }  \right)  \right)}.
\end{equation}
Here $\zeta_\pm (\eta_{\pm})$ and $\eta_{\pm}$ are given by $\eta(\beta_{\pm})$ with $\eta (\beta)$ defined by Eq.~(\ref{short-notations}) and we used
the following limiting expressions necessary to calculate the ratio (\ref{U-ratio}):
\begin{equation}
\begin{split}
& \frac{\mu^{\frac{1}{3}}_{+} g(\mu_{+})}{\mu^{\frac{1}{3}}_{-} g(\mu_{-})}
\underset{\nu \rightarrow +\infty}{\sim} \frac{\nu}{\sqrt{2}} \left( 1-\frac{1}{6 \nu^2}+O\left(\frac{1}{\nu^4} \right) \right), \qquad
\eta \underset{\beta \rightarrow 1}{\sim} \frac{2\sqrt{2}}{3}(1-\beta)^{3/2}+O \left( (1-\beta)^{5/2} \right),\\
& \frac{\phi(\zeta_{+})}{\phi(\zeta_{-})}\underset{\beta \rightarrow 1}{\sim}1-\frac{1}{10 \nu^2} +O\left(\frac{1}{\nu^4} \right), \qquad
\mu^2_{\lambda} \eta_{\lambda}\underset{\beta \rightarrow 1}{\sim} \nu^2 \eta + \lambda \frac{\sqrt{1-\beta} }{\sqrt{2}} +O \left( (1-\beta)^{3/2} \right), \\
& -\mu^{\frac{4}{3} }_{\lambda} \zeta_{\lambda}=\left(\frac{3}{2} \right)^{\frac{2}{3}} \left(\mu^2_{\lambda} \eta_{\lambda} \right)^{\frac{2}{3}}\underset{\beta \rightarrow 1}{\sim}\left(\frac{\varepsilon^2}{2} \right)^{\frac{2}{3}}  \left(1+ \lambda \frac{\sqrt{2} \sqrt{1-\beta} }{ \varepsilon^2 } +O \left( (1-\beta)^{3/2} \right)  \right).
\end{split}
\end{equation}
Since  $\gamma \underset{\beta \rightarrow 1}{\sim} \left( 1-\sqrt{2} \sqrt{1-\beta}+ (1-\beta) +O \left( (1-\beta)^{3/2} \right) \right)
$, substituting the ratio (\ref{U-ratio})
in  Eq.~(\ref{spectrum-zigzag+half})
for the spectrum in $\mathbf{K}_+$ valley we obtain 
\begin{equation}
\frac{U\left( -\frac{1}{2} \mu_{+}^2 , \sqrt{2} \beta_{+} \mu_{+} \right)}{U\left( -\frac{1}{2} \mu_{-}^2 , \sqrt{2} \beta_{-} \mu_{-} \right) } -\frac{\nu}{\sqrt{2} \gamma}\underset{\beta \rightarrow 1}{\sim} -\frac{\nu}{\sqrt{2}}  \left[ \frac{\text{Ai}' \left( -\left(\frac{\varepsilon^2}{2} \right)^{\frac{2}{3}}  \right) }{\text{Ai} \left( -\left(\frac{\varepsilon^2}{2} \right)^{\frac{2}{3}}  \right)} \left(\frac{\varepsilon^2}{2} \right)^{\frac{2}{3}}  \frac{2}{\varepsilon^2} \sqrt{2} \sqrt{1-\beta}  + \sqrt{2} \sqrt{1-\beta} \right]=0.
\end{equation}

Hence,  the spectrum for $\mathbf{K}_+$ valley
is characterized by the zeros of the following equation:
\begin{equation}
\label{Airy-K+}
\text{Ai}' \left( -\left(\frac{\varepsilon^2}{2} \right)^{\frac{2}{3}}  \right) + \left(\frac{\varepsilon^2}{2} \right)^{\frac{1}{3}} \text{Ai}\left( -\left(\frac{\varepsilon^2}{2} \right)^{\frac{2}{3}}  \right)=0.
\end{equation}
It is convenient to define in Eq.~(\ref{f-pm}) the function $f_+$
and express the spectrum  for the $\mathbf{K}_+$ valley via its zeros
$\varkappa_{+,n}$.
Then the full spectrum which includes also $\mathbf{K}_-$ valley
[see Eq.~(\ref{Airy-K-}) below] is represented by
Eq.~(\ref{e-levels-set-beta=1}).

Furthermore, under the assumption that the energies $\varepsilon_n$
are large which is certainly valid for large $n$,
Airy function and its derivative in Eq.~(\ref{f-pm})
can be expanded as follows (see Eqs.~(9.7.9) and (9.7.10) from \cite{NIST-book}):
\begin{equation}
\label{Airy-expand}
\text{Ai} \left( -x \right)\sim\frac{1}{\sqrt{\pi} x^{\frac{1}{4}}} \sin \left( \frac{2}{3} x^{3/2}+\frac{\pi}{4} \right),\quad
\text{Ai}' \left( -x \right)\sim -\frac{x^{\frac{1}{4}}}{\sqrt{\pi} } \cos \left( \frac{2}{3} x^{3/2}+\frac{\pi}{4} \right), \quad x \to \infty.
\end{equation}
Then the equation $f_{+}(-(\varepsilon^2/2)^{2/3}) =0$ for the spectrum reduces to the
trigonometric one
\begin{equation}
\label{spectrum-trig+}
-\cos \left( \frac{\varepsilon^2}{3}+\frac{\pi}{4} \right)+\sin \left( \frac{\varepsilon^2}{3} +\frac{\pi}{4} \right)=0
\end{equation}
which has the following set of the solutions
$\varepsilon_{+,n}^{\mathrm{app}}=\sqrt{3 \pi n }$ with $n=0,1,\ldots$.
This set of approximate solutions also contains the zero-energy solution, $\varepsilon_{+,0}^{\mathrm{app}}=0$,
while the full equation with the Airy function
has the lowest-energy solution $\varepsilon_{+,0}= 0.58315$.
Nevertheless, starting from $n=1$ the approximate
solutions of the trigonometric equation (\ref{spectrum-trig+})
and the full equation with the Airy function demonstrate an
excellent agreement: $\varepsilon_{+,1}^{\mathrm{app}}=3.06998$,
$\varepsilon_{+,1}=3.06965$ and
$\varepsilon_{+,2}^{\mathrm{app}}=4.34161$ and
$\varepsilon_{+,2}=4.34119$.
Thus it is enough  to omit the $n=0$ solution of Eq.~(\ref{spectrum-trig+})
for the $\mathbf{K}_+$ valley.
Accordingly, we included in the upper line
of Eq.~(\ref{e-levels-set-beta=1-approx}) only the solutions with $n=1,2, \dots$.

Similarly, for the $\mathbf{K}_-$ valley we substitute the ratio (\ref{U-ratio})
in Eq.~(\ref{spectrum-zigzag-half}) and obtain the following equation
\begin{equation}
\frac{U\left( -\frac{1}{2} \mu_{+}^2 , \sqrt{2} \beta_{+} \mu_{+} \right)}{U\left( -\frac{1}{2} \mu_{-}^2 , \sqrt{2} \beta_{-} \mu_{-} \right) } -\frac{\nu}{\sqrt{2}} \gamma \underset{\beta \rightarrow 1}{\sim} -\frac{\nu}{\sqrt{2}}  \left[ \frac{\text{Ai}' \left( -\left(\frac{\varepsilon^2}{2} \right)^{\frac{2}{3}}  \right) }{\text{Ai} \left( -\left(\frac{\varepsilon^2}{2} \right)^{\frac{2}{3}}  \right)} \left(\frac{\varepsilon^2}{2} \right)^{\frac{2}{3}}  \frac{2}{\varepsilon^2} \sqrt{2} \sqrt{1-\beta}  - \sqrt{2} \sqrt{1-\beta} \right]=0.
\end{equation}
\end{widetext}
the spectrum for $\mathbf{K}_-$ valley
is characterized by the zeros of the following equation:
\begin{equation}
\label{Airy-K-}
\begin{aligned}
\text{Ai}' \left( -\left(\frac{\varepsilon^2}{2} \right)^{\frac{2}{3}}  \right) - \left(\frac{\varepsilon^2}{2} \right)^{\frac{1}{3}} \text{Ai}\left( -\left(\frac{\varepsilon^2}{2} \right)^{\frac{2}{3}}  \right)=0
\end{aligned}
\end{equation}
Accordingly, the spectrum  for the $\mathbf{K}_-$ valley
is expressed via the zeros $\varkappa_{-,n}$ of the function $f_-$
defined by Eq.~(\ref{f-pm}).

Again using the large negative argument expansion
(\ref{Airy-expand}) applicable for the large
energies $\varepsilon_n$, one can simplify equation for
the spectrum $f_-(-(\varepsilon^2/2)^{2/3}) =0$ to the trigonometric one
\begin{equation}
\label{spectrum-trig-}
-\cos \left( \frac{\varepsilon^2}{3}+\frac{\pi}{4} \right)-\sin \left( \frac{\varepsilon^2}{3} +\frac{\pi}{4} \right)=0.
\end{equation}
Equation~(\ref{spectrum-trig-}) has the following set of the solutions
$\varepsilon_{-,n}^{\mathrm{app}}=\sqrt{3 \pi \left(n+1/2 \right)}$
with  $n=0,1,\ldots$ The comparison of
the approximate solutions given by this set and the full equation
with Airy function demonstrates a rather good agreement starting from the
$n=0$ level: $\varepsilon_{-,0}^{\mathrm{app}}=2.1708$,
$\varepsilon_{-,0}=2.15569$
and
$\varepsilon_{-,1}^{\mathrm{app}}=3.75994 $,
$\varepsilon_{-,1}=3.75805$.
Thus the whole set starting from $n=0$ can be used
in the lower line of Eq.~(\ref{e-levels-set-beta=1-approx}).
One can check that there are no collapsing solutions
$\varepsilon \sim (1-\beta^2)^{3/4}$
in the $\beta \to 1$ case.

To analyze Eqs.~(\ref{spectrum-zigzag+half}) and (\ref{spectrum-zigzag-half})
in the limit $\beta \rightarrow -1$ for the electronic levels
it is necessary to choose the large positive $\mu$
expansion (3.16) of
$U\left( -\frac{1}{2} \mu^2 , -\sqrt{2} \beta \mu \right)$
instead of employed above Eq.~(\ref{Temme-expansion}),
which converges uniformly $\beta \in [-1+\epsilon,\infty )$.
Then one can prove that the only electronic solutions
in the limit $\beta \rightarrow -1$ are collapsing, specifically,
$\varepsilon \sim (1-\beta^2)^{3/4}$.

Introducing $\nu$ defined below Eq.~(\ref{U-rewritten}),
one can rewrite Eqs.~(\ref{spectrum-zigzag+half}) and (\ref{spectrum-zigzag-half})
for $\delta = 0$ and $\beta =-1$ in the  form $f(\nu) =0$,
where 
\begin{equation}
\label{f-nu}
f(\nu) = \frac{U\left(-\frac{1}{2}-\frac{1}{2}\nu^2, -\sqrt{2} \nu \right)}{U\left(\frac{1}{2} -\frac{1}{2}\nu^2, -\sqrt{2} \nu \right)} +\frac{\nu}{\sqrt{2}}.
\end{equation}
One can check that for $\nu >0$ this equation has an infinite set
of zeros $\nu_n$:
\begin{equation}
\frac{\varepsilon}{(1-\beta^2)^{3/4}} = \nu_n.
\end{equation}
The values $\nu_{2n}$ with $n=1,2, \ldots$ correspond to the $\mathbf{K}_{+}$
valley and  $\nu_{2n+1}$ with $n=0,1, \ldots$ correspond to the
$\mathbf{K}_{-}$ valley, respectively. Thus we arrive at Eq.
(\ref{e-levels-set-beta=-1}).

\end{document}